\documentclass[10pt,reqno,fleqn,pdftex,dvips]{report}
\usepackage{amssymb,amsfonts,amsmath}
\usepackage{enumitem}
\usepackage{bm}
\usepackage{float}
\usepackage{dsfont}
\usepackage[dvips]{graphicx,color}
\usepackage{xr}

\DeclareMathOperator*{\Ind}{\mathds{1}}

\newcommand{\myABC}{ABC$\mu$}
\newcommand{\mcmcmyABC}{mcmc\myABC}

\newcommand{\bw}{\tau}

\newcommand{\Sset}{\mathbb{S}}

\newcommand{\SYMMm}{\text{SYMM}}

\newcommand{\R}{\mathbb{R}}

\newcommand{\m}{M}

\newcommand{\N}{\mathcal{N}}

\newcommand{\beps}[1][]{{\varepsilon^{#1}_{1:K}}}

\providecommand{\apostr}[1]{``#1"}		% get apostrophes around text
\providecommand{\abs}[1]{\lvert#1\rvert}	% absolute value
\providecommand{\bigabs}[1]{\bigl\lvert#1\bigr\rvert}

\DeclareMathOperator*{\me}{median}
\DeclareMathOperator*{\poi}{Poisson}
\newtheorem{theorem}{Theorem}
\newtheorem{example}[theorem]{Example}

\floatstyle{ruled}
\newfloat{boxx}{tbhp}{lop}
\floatname{boxx}{Box}

\long\def\TITLE#1{{\huge{\sc\hspace{-\parindent}#1}}}
\long\def\AUTHORS#1{\rm #1\\[3mm]}
\long\def\AFFILIATION#1#2{$^{#1}\,${\footnotesize{#2}};}

\begin{document}
\addtolength{\parskip}{0.5\baselineskip}
\bibliographystyle{plos}
    
\newenvironment{mybox}[3]{
\begin{boxx}
\colorbox{gray90}{
\begin{minipage}[b]{0.975\textwidth}
\caption{#1}\label{#2}
 #3
\end{minipage}}\end{boxx}}

\newenvironment{singlespaceddescription}{
\begin{description}
  \setlength{\itemsep}{1pt}
  \setlength{\parskip}{0pt}
  \setlength{\parsep}{0pt}
}{\end{description}}

\newenvironment{myenumerate}{
\begin{enumerate}
  \setlength{\itemsep}{1pt}
  \addtolength{\parskip}{0.5\baselineskip}
  \setlength{\parsep}{0pt}
}{\end{enumerate}}
\newenvironment{myenumerateresume}{
\begin{enumerate}[resume]
  \setlength{\itemsep}{1pt}
  \addtolength{\parskip}{0.5\baselineskip}
  \setlength{\parsep}{0pt}
}{\end{enumerate}}

\makeatletter
\newcommand\MM{\@startsection{section}{0}{0em}%
{.5\baselineskip}%
{-\fontdimen2\font
	plus -\fontdimen3\font
	minus -\fontdimen4\font
}%
{\normalfont\footnotesize\bfseries}}%
\makeatother
\renewcommand\thefootnote{\fnsymbol{footnote}}
%%%%%%%%%%%%%%%%%%%%%%%%%%%%%%
%% For titles, only capitalize the first letter
\TITLE{{\Large Notes to Robert et al.:}\\[2mm] Model criticism informs model choice and model\\[1mm] comparison}
\vspace{0.5cm}

\AUTHORS{{\bf Oliver Ratmann$^{\star,\dagger}$, Christophe Andrieu$^\ddagger$, Carsten Wiuf$^\ddagger$ and Sylvia Richardson$^\ddagger$}}
%\vspace{0.5cm}

\hspace{-\parindent}\AFFILIATION{\star}{Biology Department, Duke University, Box 90338 Durham, NC 27708, USA} \AFFILIATION{\dagger}{Statistical and Applied Mathematical Sciences Institute, Research Triangle Park, NC 27709, USA} \AFFILIATION{\ddagger}{Department of Mathematics, University of Bristol, Bristol, United Kingdom} \AFFILIATION{\ddagger}{Bioinformatics Research Center, University of Aarhus, Aarhus, Denmark} \AFFILIATION{\ddagger}{Centre for Biostatistics, Imperial College London, London, United Kingdom}

\hspace{-\parindent}Email: $^\star$ {\tt oliver.ratmann@duke.edu}
\setcounter{section}{0}
\setcounter{secnumdepth}{0}
\vspace{1cm}

In their letter to PNAS and a comprehensive set of notes on arXiv \cite{Robert2009PNAS, Robert2009b}, Christian Robert, Kerrie Mengersen and Carla Chen (RMC) represent our approach to model criticism in situations when the likelihood cannot be computed as a way to \apostr{contrast several models with each other}. In addition, guided by an analysis of scalar error terms on simple examples, RMC argue that model assessment with Approximate Bayesian Computation under model uncertainty (\myABC) is unduly challenging and question its Bayesian foundations. We thank RMC for their interest and their detailed comments on our work, which give us an opportunity to clarify the construction of \myABC\ and to explain further the utility of \myABC\ for the purpose of model criticism. Here, we provide a comprehensive set of answers to RMC's comments, which go beyond our short response \cite{Ratmann2009e}. For sake of clarity, we re-state RMC's main points in italic before we answer each of them in turn.

We wish to emphasize that the use of multiple error terms $\beps$ is a necessary and integral part of \myABC. In the first section in \cite{Ratmann2009}, we introduced \myABC\ with the number of error terms set to $K=1$ to keep the presentation simple. In retrospect, we hope that this initial simplification did not lead to confusion (although in later sections and in our applications we clearly use multiple error terms). 

%These notes are organized as follows. After a brief introduction that recapitulates the main results in Ratmann et al. \cite{Ratmann2009}, we clarify that \myABC\ is a probabilistically sound and powerful tool for criticizing a model against aspects of the observed data, and discuss further the utility of \myABC\ for the purpose of model criticism.

\subparagraph{Introduction and notation.}
Approximate Bayesian Computation (ABC) exploits model simulations $x$ of a data-generating process $\m$ for sampling from approximate posterior distributions of the model parameters $\theta$ \cite{Beaumont2002}. Typically, such predictions form the basis for model criticism \cite{Jeffreys1961}, and we propose to use the data already generated by Monte Carlo implementations of ABC for this purpose too \cite{Ratmann2009}. In \myABC, the dual use of the model predictions is reflected in an extension of the state space of the targeted random variables: whenever the simulated summaries $\Sset(x)=\big\{S_1(x),\dotsc,S_K(x)\big\}$, $x\sim f(\,\cdot\,|\theta\m)$ are sufficiently close to the observed summaries $\Sset(x_0)$, we retain not only $\theta$ but also the computed discrepancies. The rationale of \myABC\ is that small discrepancies between $x$ and the observed data $x_0$  indicate favorable $\theta$, whereas if these discrepancies are always large, the data-generating process (in short: model) $\m$ cannot describe the observed data well. The full potential of \myABC\ is realized when we compute multiple discrepancies, each for one summary statistic $S_k$, $\rho_k\big(S_k(x),S_k(x_0)\big)$. From first principles, we derived in \cite{Ratmann2009} the sampling density of the accepted pairs 
\begin{equation*}
\Big(\theta,\Big(\rho_k\big(S_k(x),S_k(x_0)\big)\Big)_{1:K}\Big),
\end{equation*}
which we denote by
\begin{equation}\label{e:jointposterior}
f_{\rho,\bw}(\theta,\beps|x_0,\m)\propto\xi_{x_0,\theta}(\beps)\pi_\theta(\theta|\m)\pi_{\beps}(\beps|\m).
\end{equation}
We obtained a formula for the \apostr{augmented likelihood} $\xi_{x_0,\theta}(\beps)$, which enables us to relate the posterior error density
\begin{equation*}
f_{\rho,\bw}(\beps|x_0,\m)=\int f_{\rho,\bw}(\theta,\beps|x_0,\m)\:d\theta
\end{equation*}
to the prior predictive error density, a well-known Bayesian quantity that was systematically discussed in a seminal paper \cite{Box1980} by Box (when $K=1$ and $\rho\big(\Sset(x),\Sset(x_0)\big)=x-x_0$). We have
\begin{equation}\label{e:posteriorerror}
f_{\rho,\bw}(\beps|x_0,\m)\propto \pi_{\beps}(\beps|\m)\:\times\:L_{\rho}(\beps|\m)
\end{equation}
where the prior predictive error density is given by
\begin{equation*}
L_{\rho}(\beps|\m)= \int\delta\big\{\big(\rho_k\big(S_k(x),S_k(x_0)\big)=\varepsilon_k\big)_{1:K}\big\}\:\pi(x|\m)\:dx,
\end{equation*}
and $\pi(x|\m)=\int f(x|\theta,\m)\pi_\theta(\theta|\m)d\theta$ denotes the prior predictive (data) density. The shorthand $\delta$ notation represents a limit of functions as detailed in Section\thinspace S1.1 of the PNAS Supplementary Material \cite{Ratmann2009}. The density $\pi_{\beps}$ is fully determined by the ABC kernel in the likelihood approximation,
\begin{equation}\label{e:approxlkl}
\begin{split}
&f_{\rho,\bw}(\theta|x_0,\m)=\int\:f_{\rho,\bw}(\theta,\beps|x_0,\m)\:d\beps\\
&\quad\propto\pi_\theta(\theta|\m)\int\pi_{\beps}\bigg(\Big(\rho_k\big(S_k(x),S_k(x_0)\big)\Big)_{1:K}\:\Big|\:\m\:\bigg)\:f(x|\theta,\m)\:dx,
\end{split}
\end{equation}
and can be interpreted as a prior density \cite{Wilkinson2008}.
The relationship Eq.\thinspace\ref{e:posteriorerror} enables us to associate a statistical interpretation to our posterior errors and to relate them to other Bayesian quantities.

\subparagraph{Standard Assumptions in ABC and \myABC.} 
We assume that (A1) $\pi_{\beps}$ factorizes into $\prod_{k=1}^K\pi_{\varepsilon_k}$, is centered at zero and only depends on a multi-dimensional scale parameter $\bw=(\bw_1,\dotsc,\bw_K)$. The main reason behind (A1) is that otherwise, the same aspects of the data might be used to adjust the ABC kernel (or \apostr{prior} density) $\pi_{\beps}$ as well as the magnitude of the errors $\rho_k\big(S_k(x),S_k(x_0)\big)$, and hence (potentially) more than once to inform our quantities of interest $f_{\rho,\bw}(\theta|x_0,\m)$ and $f_{\rho,\bw}(\beps|x_0,\m)$. Furthermore, \myABC\ might suggest to falsely reject the hypothesis that a model is an adequate representation of the data if $\pi_{\beps}$ is not centered at zero. Typical choices are $\pi_{\varepsilon_k}(\varepsilon_k|\m)=1/\bw_k\Ind\big\{\bigabs{\varepsilon_k}\leq\bw_k/2\big\}$, $\pi_{\varepsilon_k}(\varepsilon_k|\m)=(2\pi\bw_k^2)^{-1/2}\exp\big(-1/2\:\varepsilon^2/\bw_k^2\big)$ or $\pi_{\varepsilon_k}(\varepsilon_k|\m)=1/\bw_k\exp\big(-2\bigabs{\varepsilon_k}/\bw_k\big)$. We emphasize that in ABC and \myABC, (A2) the scale parameter $\bw$ of the prior $\pi_{\beps}$ is in general chosen as small as possible. Otherwise, if all model simulations are \apostr{acceptable}, we have that
\begin{equation*}
\begin{split}
&f_{\rho,\bw}(\theta|x_0,\m)\propto\pi_\theta(\theta|\m)\int \pi_{\beps}\bigg(\Big(\rho_k\big(S_k(x),S_k(x_0)\big)\Big)_{1:K}\:\Big|\:\m\:\bigg)\:f(x|\theta,\m)\:dx\\
&\quad=\pi_\theta(\theta|\m)\int \:\text{const}\:\times\:f(x|\theta,\m)\:dx\\
&\quad=\pi_\theta(\theta|\m).
\end{split}
\end{equation*}
Furthermore, (A3) the compound function $x\to\rho\big(\Sset(x),\Sset(x_0)\big)$ must be sensitive to changes in $\theta$. Otherwise, we obtain
\begin{equation*}
\begin{split}
&f_{\rho,\bw}(\theta|x_0,\m)\quad\propto\int \pi_\varepsilon\big(\:\text{const}\:\big)\:f(x|\theta,\m)\:dx\:\pi_\theta(\theta|\m)\\
&\quad=\pi_\theta(\theta|\m).
\end{split}
\end{equation*}
The idea is to construct useful discrepancies which reflect changes in the simulated data as $\theta$ changes. (A4) As in ABC, we require that $\rho_k\big(S_k(x),S_k(x_0)\big)=0$ if and only if $S_k(x)=S_k(x_0)$. In contrast to most implementations of ABC, these discrepancies should be real-valued rather than non-negative. For example, in the case of scalar summaries, we use $\rho_k\big(S_k(x),S_k(x_0)\big)= S_k(x)-S_k(x_0)$ instead of $\rho_k\big(S_k(x),S_k(x_0)\big)= \bigabs{S_k(x)-S_k(x_0)}$ \cite{Ratmann2009}. We seek to construct (A5) roughly symmetric predictive error densities $L_{\rho}(\beps|\m)$ with mode at zero under the null hypothesis that the prior model is an adequate representation of the data. Otherwise, negative small errors $\varepsilon_k\leq\bw_k$ may be significantly more (or less) frequent than positive small errors $\varepsilon_k\leq\bw_k$ under the null, and conditioning on error magnitude could result in a large negative (or positive) posterior mean error even if the prior model is correct. Finally, we assume (A6) that the cumulative density function
\begin{equation*}
\mathbb{P}_{\theta,x_0}\Big(\varepsilon_1\in\mathcal{E}_1,\dotsc,\varepsilon_K\in\mathcal{E}_K\Big)=\int_{\mathcal{X}}\Ind\Big\{\Big(\rho_k\big(S_k(x),S_k(x_{0})\big)\in\mathcal{E}_k\Big)_{1:K}\Big\}\:f(x | \theta, \m) dx
\end{equation*}
is either continuously differentiable when the observation space $\mathcal{X}$ is continuous, or a step function when $\mathcal{X}$ is finite. In this case, $\xi_{x_0,\theta}(\beps)$ can be re-written in terms of its elementary derivative. Next, in order to derive Eqns.\thinspace\ref{e:posteriorerror}-\ref{e:approxlkl}, we also assume that the data-generating process $\m$ given by $f(\,\cdot\,|\theta,\m)$ is sufficiently regular to exchange the order of integration and limits; recall Section\thinspace S1.1 of the PNAS Supplementary Material  \cite{Ratmann2009}.

\section{Construction of \myABC}
\begin{myenumerate}
\item {\it RMC point out that \apostr{the denomination [of $\xi_{x_0,\theta}(\beps)$ as a] likelihood is debatable} \cite{Robert2009b} and that \apostr{the product $\xi_{x_0,\theta}(\beps)\pi_{\beps}(\beps)$ is probabilistically incoherent} \cite{Robert2009PNAS}. This conclusion derives from at least two observations: (i) \apostr{$\xi_{x_0,\theta}$ is strictly speaking not proportional to a density in $x_0$} \cite{Robert2009b} and (ii) \apostr{$\xi_{x_0,\theta}(\beps)\pi_{\beps}(\beps)$ is not invariant under reparameterization} \cite{Robert2009b}.}% and (iii) \apostr{nonparametric $\pi_{\beps}$ may be based on the observations or on additional simulations} \cite{Robert2009b}.}

- In ABC, the observed data is reduced to a set of summary statistics and compared to simulated summaries with a positive, scalar-valued discrepancy function $\rho\big(\Sset(x),\Sset(x_0)\big)$. For the purpose of parameter inference, we only need to plug $\varepsilon=\rho\big(\Sset(x),\Sset(x_0)\big)$ into the ABC kernel. In other words, the scalar, positive error $\varepsilon$ is in ABC merely a latent random variable, introduced to facilitate Bayesian computation \cite{Andrieu2006, Andrieu2009}.

In \cite{Ratmann2009}, we derive the sampling distribution of the random variable $\varepsilon$, and recognize the utility of the related multiple error terms $\beps$, each associated to one summary, for the purpose of model criticism. To us, $\beps$ is a random variable of particular statistical interest and not any longer a latent variable introduced for computational reasons. Intuitively, we shift the observed summaries by $\beps$ and propose to infer whether summaries of $x_0$ that are shifted away from zero would occur at a higher frequency and hence be more probable than the (unshifted) observed summaries. Formally, we define and identify the probability density 
\begin{equation}\label{e:augmentedlkl}
\beps\to\xi_{\theta,x_0}(\beps)\quad=\quad\lim_{h\to 0}\int \delta_h\Big(\Big(\rho_k\big(S_k(x),S_k(x_0)\big)-\varepsilon_k\Big)_{1:K}\Big)\:f(x|\theta,\m)\:dx,
\end{equation}
where the $\delta_h$ function is given in Section\thinspace1.1 of the PNAS Supplementary Material \cite{Ratmann2009}. For any given $x_0$ and $\theta$, $\xi_{\theta,x_0}(\beps)$ is the infinitesimal frequency with which we observe the multi-dimensional error $\beps$. As RMC remark insightfully, it can be called a predictive error density that conditions on the observed data and the model parameter $\theta$. In \cite{Ratmann2009}, we termed 
\begin{equation}\label{e:augmentedlkl2}
\theta,\beps\to f_{\rho,\bw}(x_0|\theta,\beps)\quad=\quad\xi_{\theta,x_0}(\beps)
\end{equation}
an \apostr{augmented likelihood} simply to indicate that the state space was extended. 
%Motivated by nascent delta functions (i.e. $f(z)=\lim_{h\to 0}\int\delta_h(x-z)\:f(x)\:dx$), we interpret Eq.\thinspace\ref{e:augmentedlkl} as an \apostr{augmented likelihood} function $(\theta,\beps)\to f_{\rho,\bw}(x_0|\theta,\beps,\m)$. In \cite{Ratmann2009}, we do not call $\xi_{\theta,x_0}(\beps)$ a \apostr{likelihood}, and $\xi_{\theta,x_0}(\beps)$ does not correspond to the probability of the data under a particular sampling model.

\begin{example}\label{ex:poisssonauglkl}
Suppose we observe a single, one-dimensional data point $x_0$, and let us believe it is Poisson distributed with rate $\theta$ (denoted by $\m_1$). Consider the scalar error $\varepsilon=x-x_0$. By construction, we have 
\begin{equation*}
\xi_{\theta,x_0}(\varepsilon)\quad=\quad\lim_{h\to 0}\int \delta_h\big(x-x_0=\varepsilon\big)\:\poi(x;\theta)\:dx\quad=\quad\frac{\theta^{x_0+\varepsilon}e^{-\theta}}{(x_0+\varepsilon)!}\Ind\big\{x_0+\varepsilon\in [0,\infty)\big\},
\end{equation*}
and the right hand side equals in $\varepsilon$ a Poisson distribution shifted by $-x_0$ and in $x_0$ a Poisson distribution shifted by $-\varepsilon$. Thus, when interpreted as a function in $x_0$, $\xi_{\theta,x_0}(\varepsilon)$ is also defined for negative values. 

RMC's illuminating Poisson example serves to demonstrate how $\xi_{\theta,x_0}(\varepsilon)$ differs from a \apostr{likelihood}. However, RMC go beyond our construction Eq.\thinspace\ref{e:augmentedlkl}  and truncate $x_0\to\xi_{\theta,x_0}(\varepsilon)$ to positive values so as to re-adjust $\xi_{\theta,x_0}(\varepsilon)$ to the likelihood $f(x_0|\theta,\m_1)$ that is only defined for positive $x_0$ \cite{Robert2009PNAS, Robert2009b}. To be clear, this re-adjustment is not part of \myABC.
\end{example}

Eq.\thinspace\ref{e:augmentedlkl} corresponds to a non-parametric evaluation of the sampling model in the context of model uncertainty. We adhere to the sampling model in that data is simulated under the likelihood, $x\sim f(\,\cdot\,|\theta,\m)$, and probe the model predictions in several directions at the same time. If $\beps=0$, we have with (A4) that $\xi_{\theta,x_0}(\beps)$ corresponds to the probability of the observed summaries under $\theta$. For error terms different from zero, we quantify the probability of deviations from the observed summaries under the sampling model. Labeling $f_{\rho,\bw}(x_0|\theta,\beps)$ Eq.\thinspace\ref{e:augmentedlkl2} a \apostr{shifted likelihood} seems therefore more appropriate. Because we only shift the observed summaries in Eq.\thinspace\ref{e:augmentedlkl} (with no further re-adjustments towards the original likelihood as in  \cite{Robert2009PNAS, Robert2009b}), the re-normalization required when considering Eq.\thinspace\ref{e:augmentedlkl2} as a function in $x_0$ does not depend on $\beps$, and $f_{\rho,\bw}(x_0|\theta,\beps)$ is proportional to a density in $x_0$.

Next, let us recall that our error terms $\varepsilon_k$ correspond directly to the compound functions $x\to\rho_k\big(S_k(x),S_k(x_0)\big)$. Therefore, in \myABC, a transformation of $\varepsilon_k$ implies a change in how the data is being summarized. Typically, such a change requires to modify the scale parameter $\bw$ of the prior density $\pi_{\beps}$ when the scale of the discrepancies changes too. Therefore, transformations of the product $\xi_{x_0,\theta}(\varepsilon)\pi_\varepsilon(\varepsilon)$ must also change $\bw$ in $\pi_{\beps}$ when the Jacobian is not constant. 

In the ABC literature, it is well-known that the approximate posterior density $f_{\rho,\bw}(\theta|x_0,\m)$ depends on the choice of discrepancies and the stringency of $\bw$ \cite{Beaumont2002, Pritchard1999, Joyce2008}. Since \myABC\ only uses the information provided in ABC to a fuller extent, the joint posterior density $f_{\rho,\bw}(\theta,\beps|x_0,\m)$ is equally sensitive to changes in the compound functions $x\to\rho_k\big(S_k(x),S_k(x_0)\big)$ and the vector $\bw$. In other words, the ABC and \myABC\ target densities $f_{\rho,\bw}(\theta|x_0,\m)$ and $f_{\rho,\bw}(\theta,\beps|x_0,\m)$ are not invariant under different approximation schemes. This leaves \myABC\ probabilistically sound, but warrants particular caution and calls for sensitivity analyses, perhaps to a larger extent than is common practice.
\end{myenumerate}

\section{Model assessment}
\begin{myenumerateresume}
\item {\it Model assessment with \myABC\ requires that \apostr{the data is informative} and \apostr{is challenging} \cite{Robert2009PNAS, Robert2009b}. In the location-family example \cite{Robert2009b} it is shown that the posterior error equals the prior error if the prior predictive density is flat.}

- We agree with RMC that \myABC\ cannot criticize a model when the observed data $x_0$ reduces to a single, one-dimensional data point, as in their examples \cite{Robert2009b} on page 1-2. More generally, we showed that the posterior error can be interpreted as a weighted prior predictive error, Eq.\thinspace\ref{e:posteriorerror} \cite{Ratmann2009}. If the prior predictive error is uninformative, then the posterior error will simply reflect the weighting.

\begin{example}\label{ex:gauss_uniformpriorpredictive}
Suppose a Gaussian likelihood model $\m_2$ with unknown mean $\theta$ and fixed variance $1$, and a Gaussian prior density $\pi_\theta(\theta|\m_2)=\N(\theta;\theta^\star,h^2)$. We have
\begin{equation*}
\begin{split}
&\pi(x|\m_2)=\int\:\N(x;\theta,1)\:\N(\theta;\theta^\star,h^2)\:d\theta\\
&\quad=\N(x;\theta^\star,h^2+1),
\end{split}
\end{equation*}
and $L_{\rho,\bw}(\varepsilon|\m_2)=\N(\varepsilon;\theta^\star-x_0,h^2+1)$. We mimic a situation where $\pi_\theta$ is uniform by letting $h\to\infty$, so that $\pi(x|\m_2)$ and $L_{\rho,\bw}(\varepsilon|\m_2)$ become improper. Suppose further a Gaussian error density $\pi_{\varepsilon}(\varepsilon|\m_2)=\N(\varepsilon;0,\bw^2)$. Then, $f_{\rho,\bw}(\varepsilon|x_0,\m_2)=\N(\varepsilon;0,\bw^2)$. 

Likewise, when models have comparable parameter spaces, then the Bayes' factor will be indecisive under non-informative priors $\pi_\theta$ \cite{Berger2001}. Consider the alternative Gaussian model $\m_2^\prime$ defined by $f(x|\theta,\m_2^\prime)=\N(x;\theta,3)$, the same prior density $\pi_\theta$, and let us focus on the approximate Bayes' factor
\begin{equation*}
B_{\rho,\bw}\quad=\quad\frac{f_{\rho,\bw}(x_0|\m_2^\prime)}{f_{\rho,\bw}(x_0|\m_2)}\quad=\quad\bigg(\int f_{\rho,\bw}(x_0|\theta,\m_2^\prime)\pi_\theta(\theta|\m_2^\prime)\:d\theta\bigg)\Big/\bigg(\int f_{\rho,\bw}(x_0|\theta,\m_2)\pi_\theta(\theta|\m_2)\:d\theta\bigg)
\end{equation*}
to mimic the situation that we cannot readily evaluate the likelihood. We obtain
\begin{equation*}
B_{\rho,\bw}=\sqrt{\frac{\bw^2+h^2+1}{\bw^2+h^2+3}}\exp\bigg(\frac{\overline{x}^2_0}{(\bw^2+h^2+1)(\bw^2+h^2+3)}\bigg),
\end{equation*}
which tends rapidly to one as $h\to\infty$. 

In this setting, both our posterior error and approximate Bayes' factor give reasonable answers for the purpose of model criticism and model comparison respectively. Based on one data point, we cannot reject the current model $\m_2$ and likewise, the approximate Bayes' factor for choosing among $\m_2$ and a comparable model is indecisive. 
%Let us also note that, as $h\to\infty$, the marginal likelihood $\N(x_0;\theta^\star,h^2+1)$ tends to zero. Considering a similar situation for an alternative model, the Bayes' factor reflects prior choice and becomes indeterminate.
\end{example}

Clearly, there is no guarantee that \myABC\ always uncovers existing model mismatch. But how difficult is it to uncover existing discrepancies with \myABC\ in practice? Typically, $x_0$ contains some structure and/or repeated observations. Instead of using just one data point in Example\thinspace\ref{ex:gauss_uniformpriorpredictive}, let us imagine a data set of $100$ samples and summarize this data with two statistics, leading to a two-dimensional posterior error density.

\begin{example}\label{ex:gauss_uniformpriorpredictive2}
Consider a data set $x_0$ of $100$ independent samples that are Exponentially distributed with rate $1/\mu_t=0.2$. We believe that each sample is generated from $\N(\,\cdot\,;\theta,1)$ and consider a Gaussian prior density $\pi_\theta(\theta|\m_2)=\N(\theta;\theta^\star,h^2)$. We summarize the data with the sample mean $\overline{x}_0$ and the sample median $\me(x)$, use the discrepancies $\rho\big(S_k(x),S_k(x_0)\big)=S_k(x)-S_k(x_0)$ and consider the prior density $\pi_{\beps}(\beps|\m_2)=\prod_k1/\bw_k\exp\big(-2\abs{\varepsilon_k}/\bw_k\big)$ with $\bw_k=0.1$.

To illustrate that \myABC\ may reveal inappropriate prior specifications, we set $\theta^\star=0$ and $h^2=0.1$. We applied the Metropolis-Hastings sampler proposed by Marjoram et al. \cite{Marjoram2003} and recorded the computed discrepancies to estimate our posterior error (\mcmcmyABC\ see page\thinspace\pageref{s:mcmcabcmu}). A more detailed discussion of various algorithms to sample from the \myABC\ target density Eq.\thinspace\ref{e:jointposterior} will appear elsewhere. Figures\thinspace\ref{f:meanmedian}A-B show that the marginal densities $f_{\rho,\bw}(\varepsilon_k|x_0,\m_2)$ are far from zero, suggesting that our strong prior beliefs are inadequate to explain the data.

To illustrate that \myABC\ may identify a faulty sampling model, we set $\theta^\star=5$, $h^2=100000$. Again, we estimated the \myABC\ target density numerically with \mcmcmyABC. Even though $\pi_\theta$ is essentially flat, our marginal posterior errors do not center at zero, see Figures\thinspace\ref{f:meanmedian}C-D.

In \cite{Ratmann2009}, we investigated primarily the marginal posterior densities $f_{\rho,\bw}(\varepsilon_k|x_0,\m)=\int f_{\rho,\bw}(\beps|x_0,\m)\:d\varepsilon_{-k}$. Here, we also show heat plots which reflect more comprehensively the multi-dimensional character of our error density $f_{\rho,\bw}(\varepsilon_{\overline{x}},\varepsilon_{\me}|x_0,\m_2)$ in Figure\thinspace\ref{f:meanmedian1b}A-B.
\end{example} 
\begin{figure}[tbp]
\vspace{-1cm}
\begin{minipage}[b]{\textwidth}
		\begin{minipage}[b]{0.24\textwidth}
			\begin{minipage}[b]{0.001\textwidth}
				{\bf A}\newline\vspace{-1cm}

			\end{minipage}
			\begin{minipage}[b]{0.99\textwidth}
				\includegraphics[type=pdf,ext=.pdf,read=.pdf,width=\textwidth]{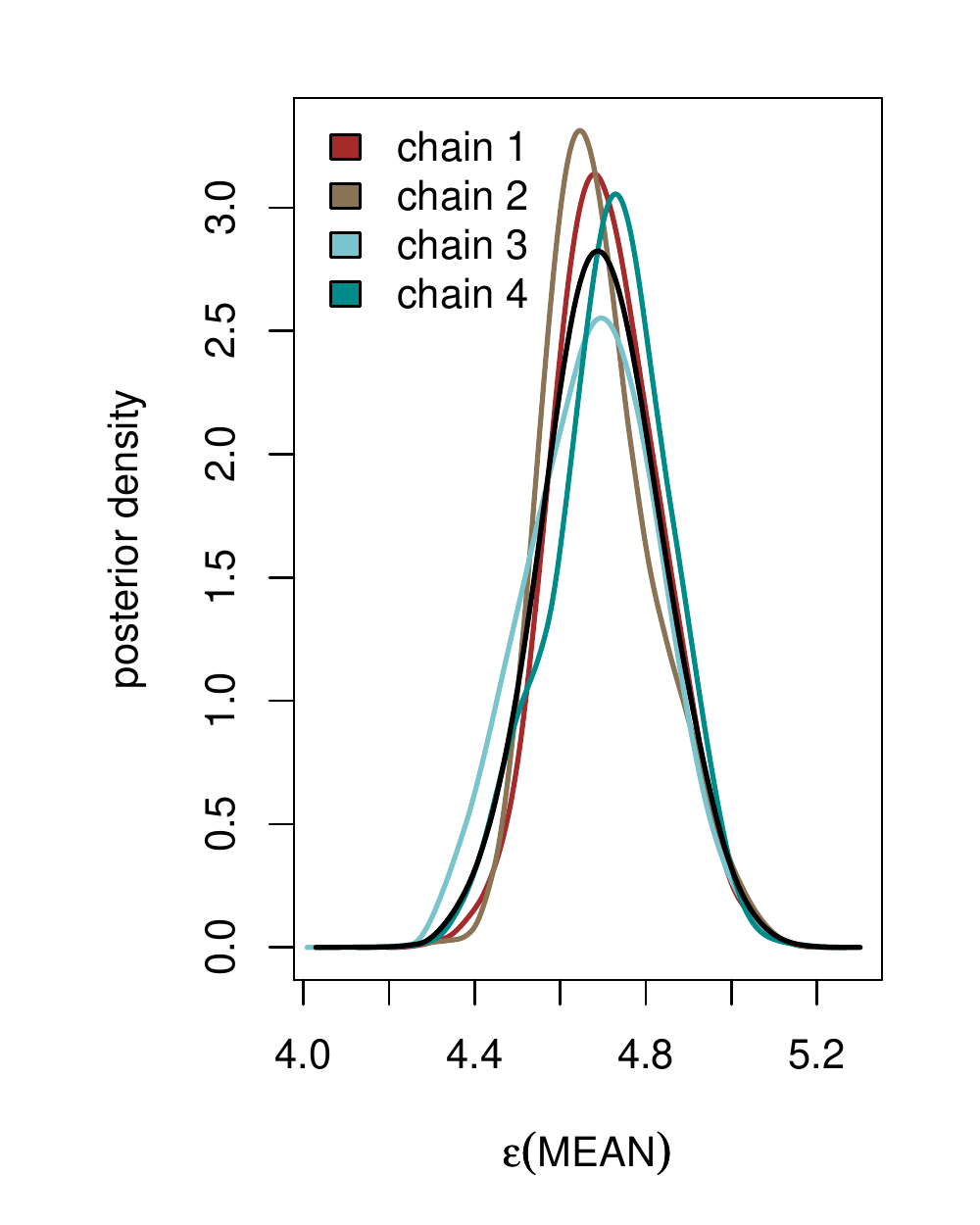}
			\end{minipage}
		\end{minipage}
		\begin{minipage}[b]{0.24\textwidth}
			\begin{minipage}[b]{0.001\textwidth}
				{\bf B}\newline\vspace{-1cm}

			\end{minipage}
			\begin{minipage}[b]{0.99\textwidth}
				\includegraphics[type=pdf,ext=.pdf,read=.pdf,width=\textwidth]{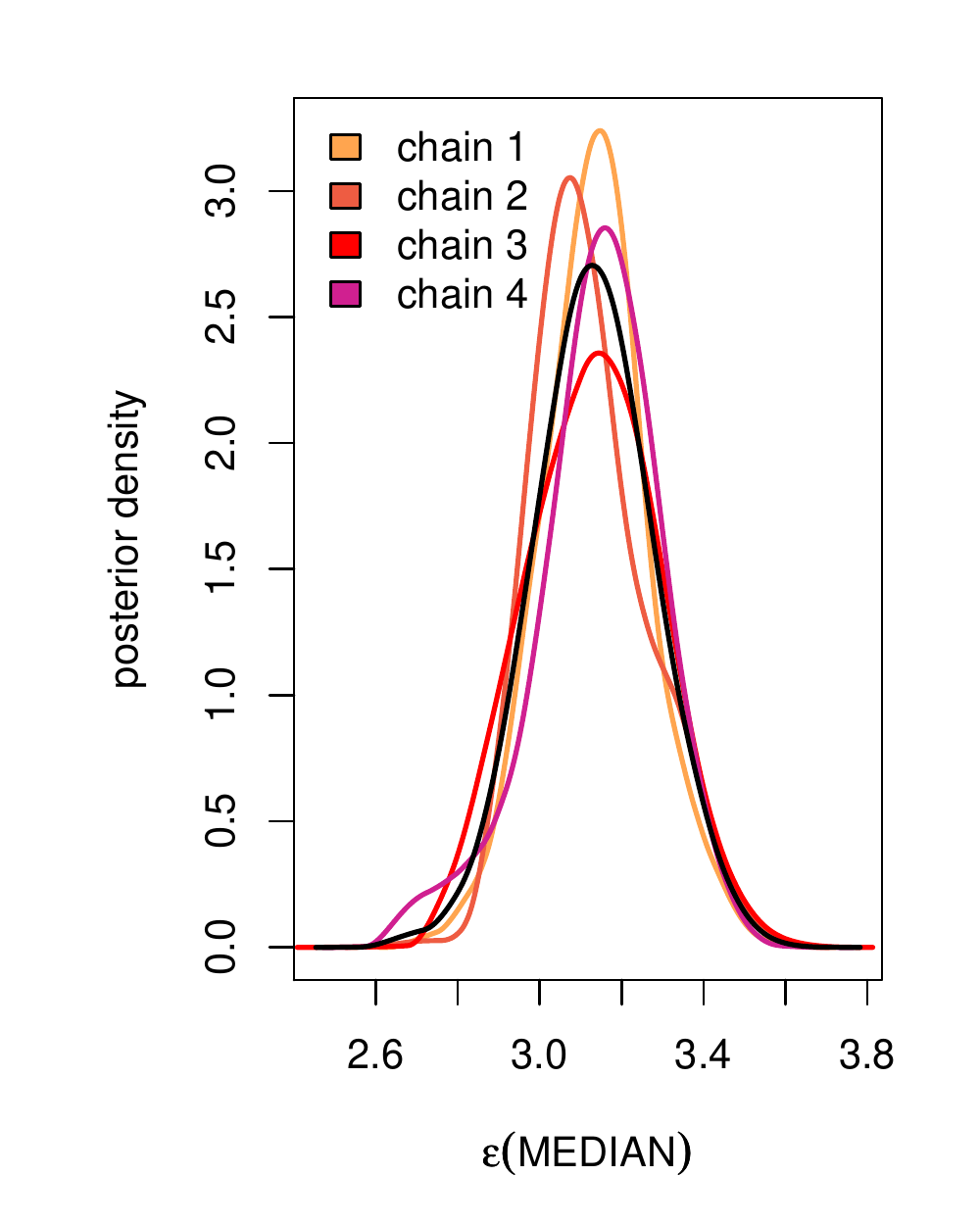}
			\end{minipage}
		\end{minipage}
		\begin{minipage}[b]{0.24\textwidth}
			\begin{minipage}[b]{0.001\textwidth}
				{\bf C}\newline\vspace{-1cm}

			\end{minipage}
			\begin{minipage}[b]{0.99\textwidth}
				\includegraphics[type=pdf,ext=.pdf,read=.pdf,width=\textwidth]{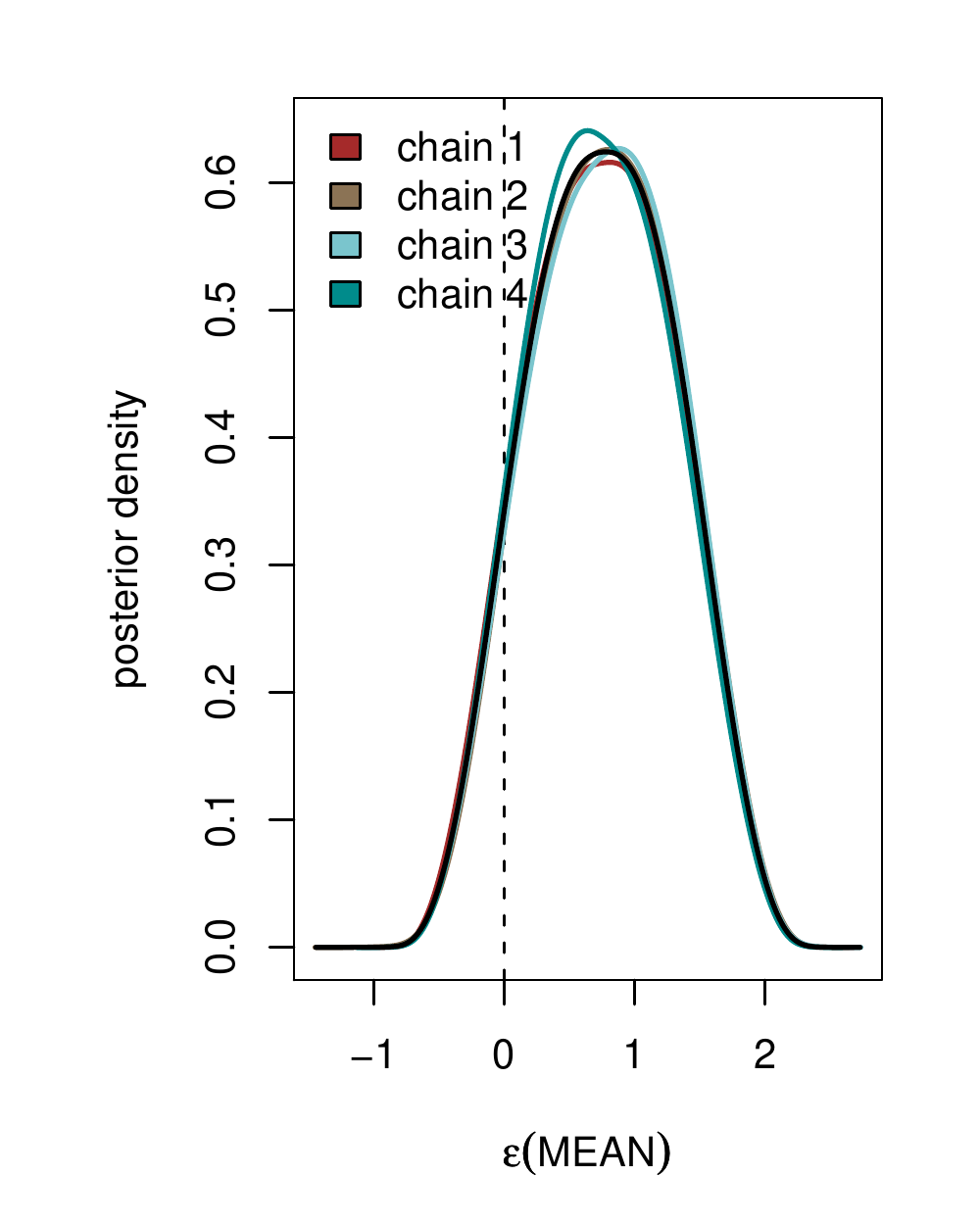}
			\end{minipage}
		\end{minipage}
		\begin{minipage}[b]{0.24\textwidth}
			\begin{minipage}[b]{0.001\textwidth}
				{\bf D}\newline\vspace{-1cm}

			\end{minipage}
			\begin{minipage}[b]{0.99\textwidth}
				\includegraphics[type=pdf,ext=.pdf,read=.pdf,width=\textwidth]{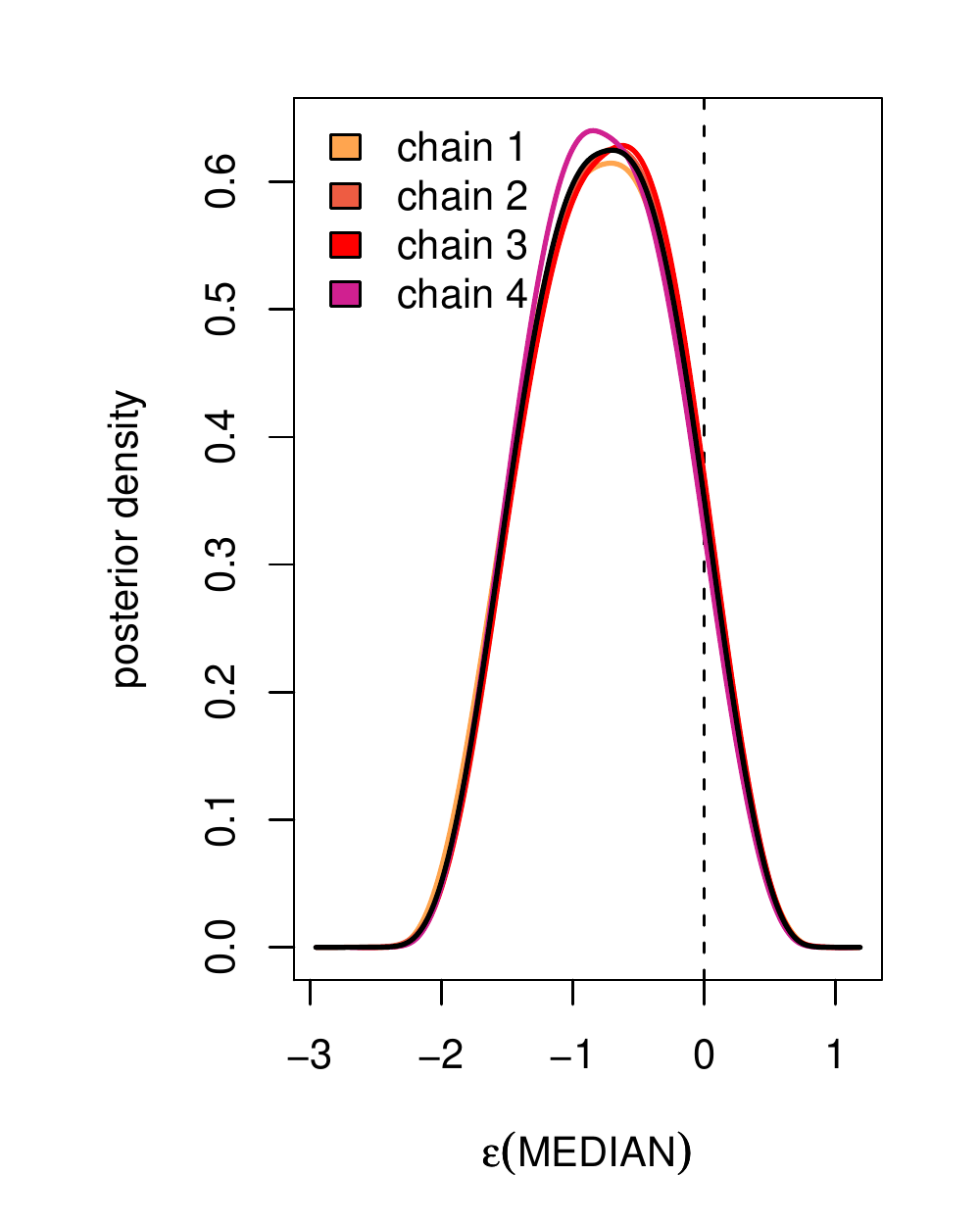}
			\end{minipage}
		\end{minipage}
	\end{minipage}
	\caption{Numerical reconstructions of the densities $f_{\rho,\bw}(\varepsilon_{\overline{x}}|x_0,\m_2)$, $f_{\rho,\bw}(\varepsilon_{\me}|x_0,\m_2)$ in Example\thinspace\ref{ex:gauss_uniformpriorpredictive2}, obtained with samples generated by \mcmcmyABC. The (A-B) posterior error under inappropriate prior specifications $\pi_\theta$ and the (C-D) posterior error under essentially flat prior specifications $\pi_\theta$ suggest model mismatch.}\label{f:meanmedian}
\end{figure}

\begin{figure}[tbp]
\vspace{0.5cm}
\begin{minipage}[b]{\textwidth}
	\centering
		\begin{minipage}[b]{0.4\textwidth}
			\begin{minipage}[b]{0.001\textwidth}
				{\bf A}\newline\vspace{6.5cm}

			\end{minipage}
			\begin{minipage}[b]{0.99\textwidth}
				\includegraphics[type=pdf,ext=.pdf,read=.pdf,width=\textwidth]{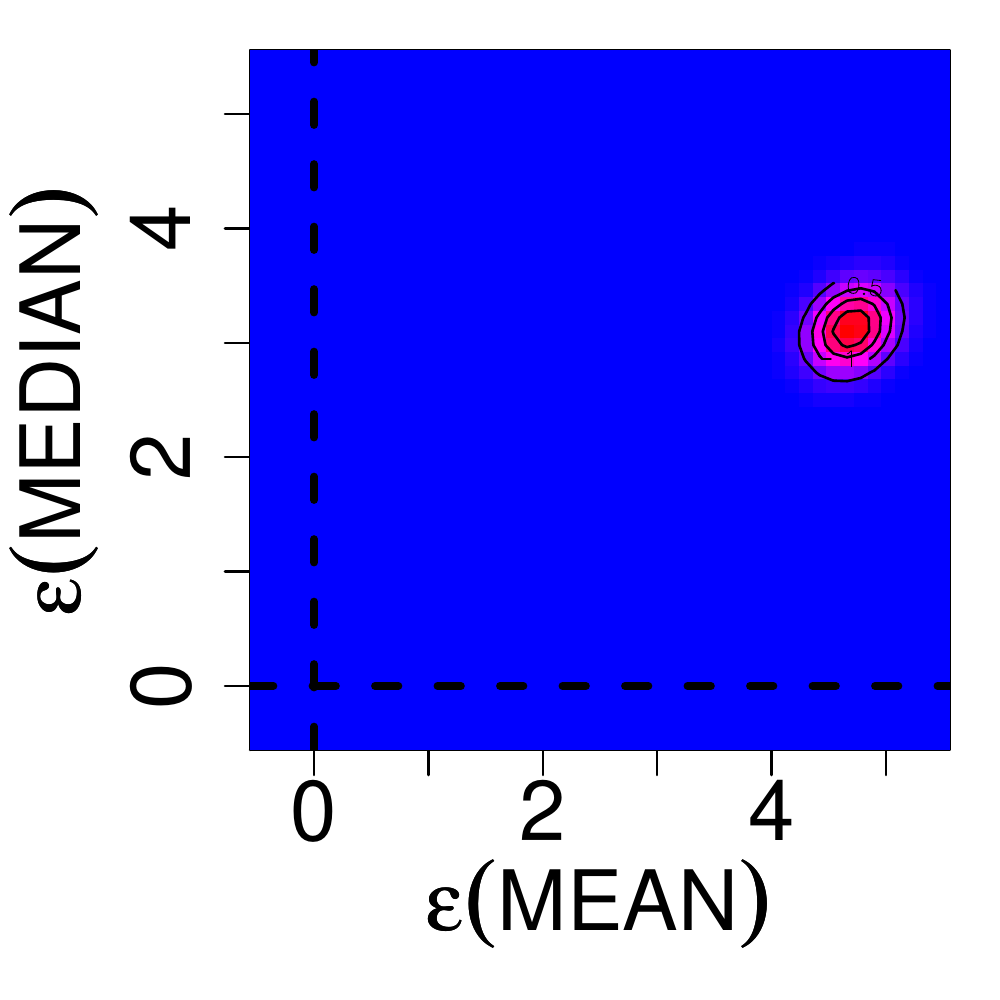}
			\end{minipage}
		\end{minipage}
		\hspace{1.5cm}
		\begin{minipage}[b]{0.4\textwidth}
			\begin{minipage}[b]{0.001\textwidth}
				{\bf B}\newline\vspace{6.5cm}

			\end{minipage}
			\begin{minipage}[b]{0.99\textwidth}
				\includegraphics[type=pdf,ext=.pdf,read=.pdf,width=\textwidth]{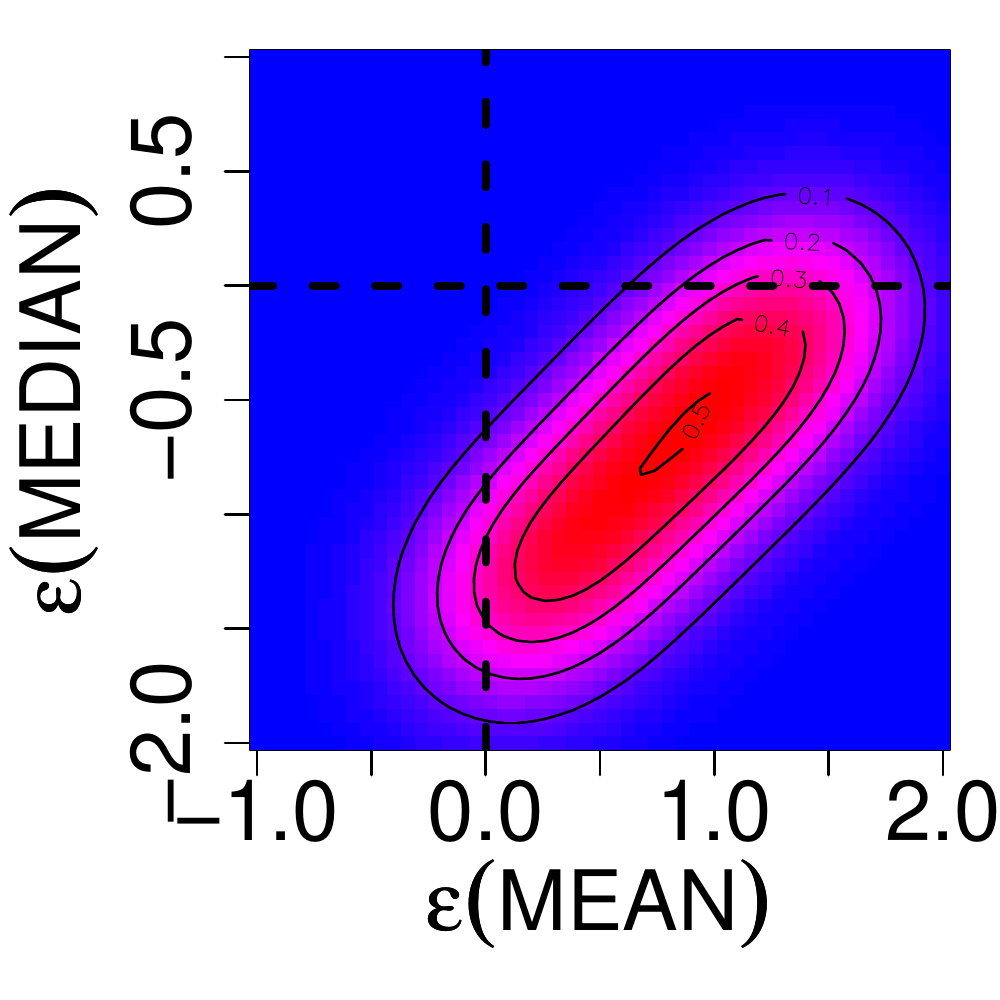}
			\end{minipage}
		\end{minipage}
	\end{minipage}
	\caption{Heat plots of the density $f_{\rho,\bw}(\varepsilon_{\overline{x}},\varepsilon_{\me}|x_0,\m_2)$ in Example\thinspace\ref{ex:gauss_uniformpriorpredictive2}, obtained with samples generated by \mcmcmyABC, (A) under inappropriate prior specifications $\pi_\theta$ and (B) under essentially flat prior specifications $\pi_\theta$.}\label{f:meanmedian1b}
\end{figure}

Intuitively, \myABC\ will indicate model mismatch whenever all discrepancies are simultaneously not close to zero for any $\theta$. To escape unidentifiability, the crux in Example\thinspace\ref{ex:gauss_uniformpriorpredictive2} is to use multiple error terms associated to co-dependent summaries that may reveal model inconsistencies, see also \cite{Ratmann2009} for a similar example. In real-world applications, (most) summary statistics are usually co-dependent, rendering \myABC\ a potentially very powerful method to reveal model inconsistencies. Because model inconsistencies can only increase with the inclusion of new summary statistics, we are typically prepared to use a large set of summaries. Moreover, it is not required that these co-dependent summaries are sufficient for $\theta$ under the data-generating process $\m$, as we illustrate in \cite{Ratmann2009e}, Figure\thinspace1. A more detailed discussion will appear elsewhere; here we only note that these properties are appealing because in real-world applications of ABC and \myABC, it is typically not known whether any set of summaries is sufficient for the parameters of a given model while it is relatively easy to come up with co-dependent summaries.

However, the extent to which $f_{\rho,\bw}(\beps|x_0\m)$ merely reflects the weighting $\pi_{\beps}$ should be checked, because the discrepancies might not retain enough information of the data to question a model (recall A3). The perhaps simplest (but not necessarily successful) approach is to compare the posterior error density to the shape of $\pi_{\beps}$. Reassuringly, in our real-world applications, $f_{\rho,\bw}(\beps|x_0\m)$ differs markedly from $\pi_{\beps}$; see e.g. Figure 3 in the PNAS paper where the prior error density is indicated in dotted lines. Crucially, since \myABC\ does not reject a model when $f_{\rho,\bw}(\beps|x_0\m)$ is close to $\pi_{\beps}$ (recall A1), no harm is done should the discrepancies not be informative.

The power of \myABC\ in assessing goodness-of-fit stems, firstly, from probing a model in multiple directions at the same time. We hope that our simple examples illuminate the contribution of model inconsistencies, as reflected in multiple error terms, to model assessment. Secondly, \myABC\ makes possible to criticize a model whose likelihood cannot be readily evaluated, and does not incur any extra computational cost when compared to ABC (in contrast to related predictive approaches discussed in point 4 below). 
\end{myenumerateresume}

\section{Model criticism}
\begin{myenumerateresume}
\item {\it \myABC \apostr{is strongly impacted by prior modeling [and] fails to condition on the observed data} \cite{Robert2009PNAS}.}

- The ABC kernel, which can be interpreted as a prior density $\pi_{\beps}$ \cite{Wilkinson2008}, is at the heart of ABC (recall A1) and modulates the degree to which ABC and \myABC\ condition on the observed data. In other words, the ABC and \myABC\ target densities are sensitive to the choice of $\pi_{\beps}$ and particularly its scale parameter $\bw$. Indeed, \myABC\ conditions on the observed data by accepting $\theta$ in relation to the magnitudes of the $K$ computed discrepancies taken together. Accordingly, under (A2, A3), the posterior error density $f_{\rho,\bw}({\beps}|x_0,\m)$ updates the prior predictive error density $L_{\rho}(\beps|\m)$; see Example\thinspace\ref{ex:gauss_criticizefittedmodel} below. Based on the observation that small error boosts the weight of the associated value of $\theta$ that are simulated from $\pi_\theta$, we say that \apostr{\myABC\ criticizes a fitted model}. This can be illustrated with the location family in Example\thinspace\ref{ex:gauss_uniformpriorpredictive}.

\begin{example}\label{ex:gauss_criticizefittedmodel}
Consider again the Gaussian likelihood model $\m_2$ and a Gaussian prior density $\pi_\theta$ as in Example\thinspace\ref{ex:gauss_uniformpriorpredictive}. For our illustration purposes, let us choose $\pi_\theta$ broad but not flat: $h^2=9$. In this case, 
\begin{equation*}
\begin{split}
&f_{\rho,\bw}(\varepsilon|x_0,\m_2)\propto\N(\varepsilon;\theta^\star-x_0,10)\N(\varepsilon;0,\bw^2)\\
&\quad=\N(\varepsilon;\tilde{\theta},\tilde{\sigma}^2)
\end{split}
\end{equation*}
where $\tilde{\theta}= \big[\tau^2\big/(\tau^2+10)\big]\times\big(\theta^\star-x_0\big)$ and $\tilde{\sigma}^2=10\tau^2/(\tau^2+10)\leq 10$. The posterior error density \apostr{updates} the prior predictive error in that the variance of $f_{\rho,\bw}(\varepsilon|x_0,\m_2)$ is smaller than the one of $L_{\rho,\bw}(\varepsilon|\m_2)$. Furthermore, we observe that $\bigabs{\tilde{\theta}}$ is smaller than the absolute mean of $L_{\rho,\bw}(\varepsilon|\m_2)$, reflecting the fact that $f_{\rho,\bw}(\varepsilon|x_0,\m_2)$ criticizes a fitted model rather than the prior model $(\m_2,\pi_\theta)$.
\end{example}

For the purpose of model criticism, it is important to recognize that the dependency of our posterior error on $\pi_{\beps}$ is a good thing to the extent to which the prior $\pi_\theta$ is not an adequate model parameterization. The smaller $\bw$ can be chosen, the more we are able to criticize a fitted model and the more we attenuate the influence of $\pi_\theta$ in $f_{\rho,\bw}(\beps|x_0,\m)$. The latter point can also be illustrated with the location family in Example\thinspace\ref{ex:gauss_uniformpriorpredictive}, recall Section\thinspace S1.2 in the PNAS Supplementary Material. Let us illustrate the influence of $\pi_\theta$ and $\pi_{\beps}$ when the summaries are co-dependent but not sufficient for $\theta$ under model $\m$.

\begin{figure}[tbp]
\vspace{0.5cm}

\begin{minipage}[b]{\textwidth}
		\begin{minipage}[b]{0.32\textwidth}
			\begin{minipage}[b]{0.001\textwidth}
				{\bf A}\newline\vspace{5cm}

			\end{minipage}
			\begin{minipage}[b]{0.99\textwidth}
				\includegraphics[type=pdf,ext=.pdf,read=.pdf,width=\textwidth]{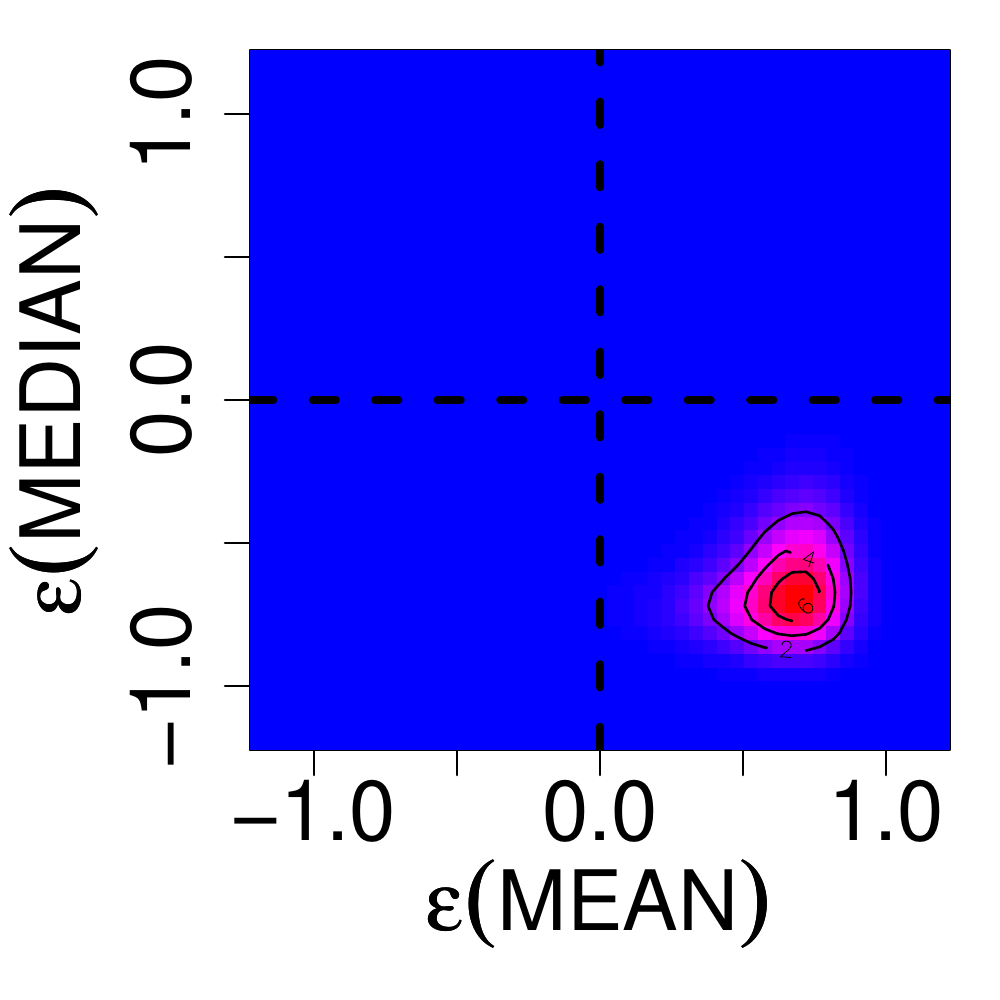}
			\end{minipage}
		\end{minipage}
		\begin{minipage}[b]{0.32\textwidth}
			\begin{minipage}[b]{0.001\textwidth}
				{\bf B}\newline\vspace{5cm}

			\end{minipage}
			\begin{minipage}[b]{0.99\textwidth}
				\includegraphics[type=pdf,ext=.pdf,read=.pdf,width=\textwidth]{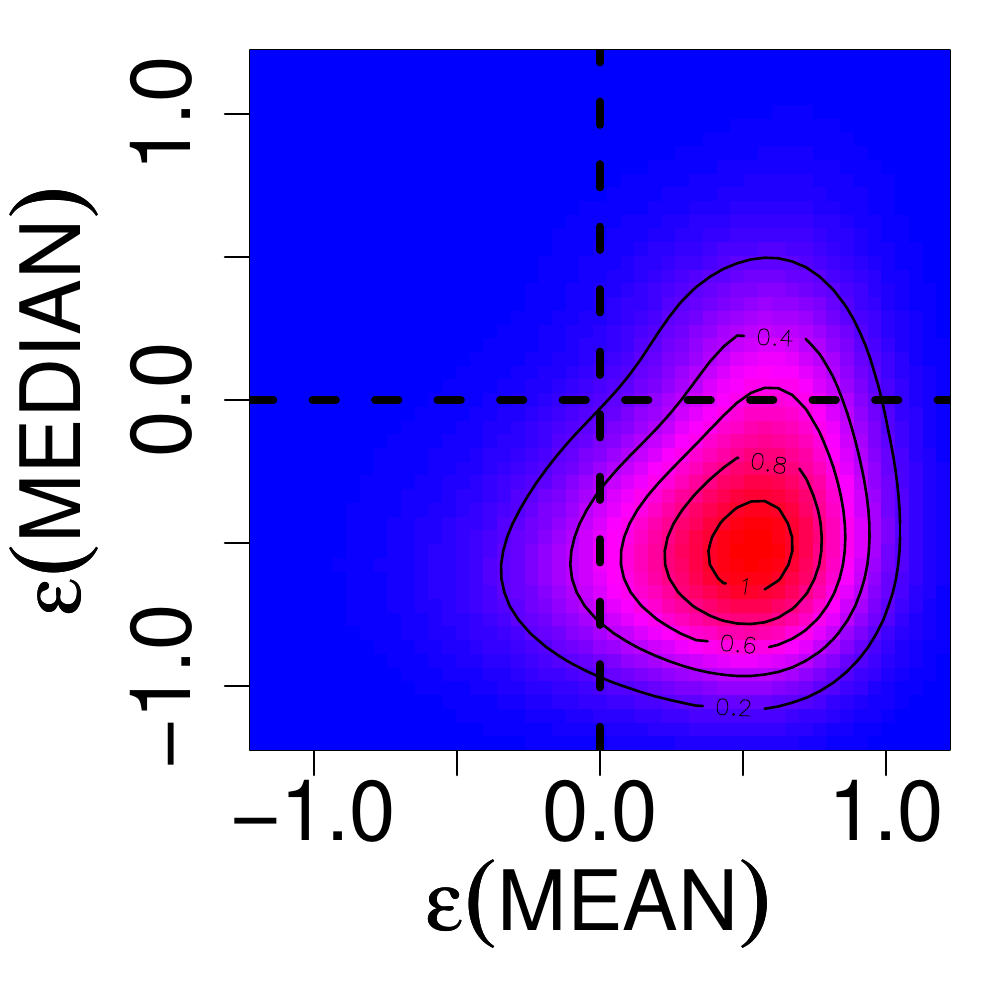}
			\end{minipage}
		\end{minipage}
		\begin{minipage}[b]{0.32\textwidth}
			\begin{minipage}[b]{0.001\textwidth}
				{\bf C}\newline\vspace{5cm}

			\end{minipage}
			\begin{minipage}[b]{0.99\textwidth}
				\includegraphics[type=pdf,ext=.pdf,read=.pdf,width=\textwidth]{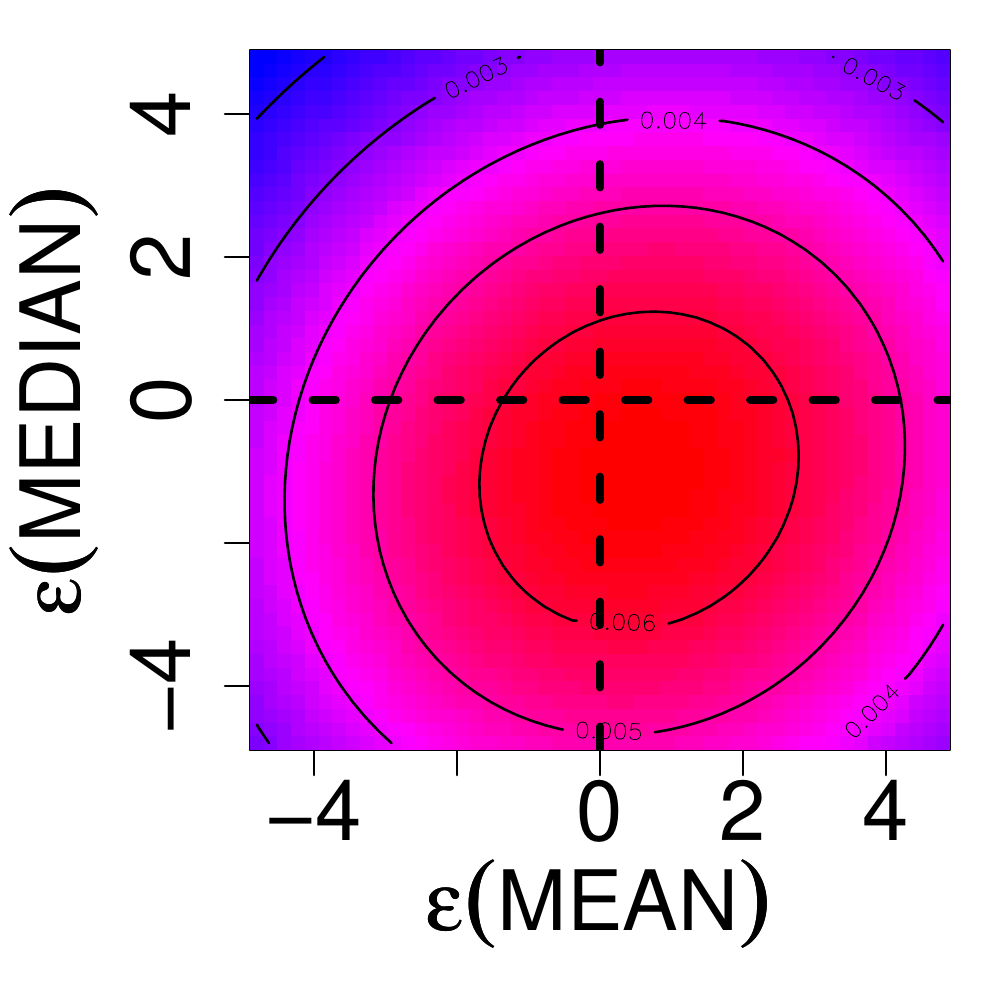}
			\end{minipage}
		\end{minipage}
		
	\end{minipage}
	\caption{Heat plots of our posterior error density $f_{\rho,\bw}(\varepsilon_{\overline{x}},\varepsilon_{\text{median}}|x_0,\m_3)$ in Example\thinspace\ref{s:gaussmusigmapriors} for broad $\pi_\theta$ (A) $\mu_0=1000$, (B) $\alpha_0=2$ and $\beta_0=\bw_0=1000$ and when $\bw$ is set too large, (C) $\alpha_0=2$ and $\beta_0=\bw_0=1000$ and $\bw_k=6.4$.}\label{f:meanmedian2}
\end{figure}

\begin{example}\label{s:gaussmusigmapriors}
Consider again the data set $x_0$ of $100$ independent samples that are Exponentially distributed with rate $0.2$, suppose now that each sample is generated according to a Gaussian likelihood model with unknown mean $\mu\in\R$ and $\sigma^2\geq 0$ (denoted by $\m_3$) and summarize the data with the sample mean and median. We consider $\rho_k(S_k(x),S_k(x_0))=S_k(x)-S_k(x_0)$, $\pi_{\beps}(\beps|\m_3)=\prod_{k=1}^K1/\bw_k\Ind\big\{\bigabs{\varepsilon_k}\leq\bw_k/2\big\}$, and the prior density $\pi_\theta(\theta|\m_3)=\pi(\mu|\m_3)\pi(\sigma^2|\m_3)$ where 
\begin{equation*}
\begin{split}
&\pi(\sigma^2|\m_3)= \text{IG}(\sigma^2;\alpha_0,\beta_0)\\
&\pi(\mu|\m_3)\propto\Ind\big\{\bigabs{\mu-\mu_0}\leq\bw_0\big\}.
\end{split}
\end{equation*}
In \cite{Ratmann2009e}, Figure\thinspace1, we chose a slightly different prior density $\pi_{\theta}$ with hyperparameters $\mu_0=5$, $\bw_0=10$, $\alpha_0=4$ and $\beta_0=75$ such that the prior means of $\mu$ and $\sigma^2$ are $5$ and $\beta_0/\big(\alpha_0-1\big)=25$, matching the empirical mean and the standard deviation of the observed data. 

To illustrate that \myABC\ uncovers existing model mismatch with co-dependent summaries that are not sufficient for $\theta$ under $\m_3$ even when $\pi_{\theta}$ differs markedly from the data or is uninformative, we now vary these hyperparameters. First, let us set $\bw_0=1000$. We ran \mcmcmyABC\ (see page\thinspace\pageref{s:mcmcabcmu}) for 10$,$000 iterations to sample from $f_{\rho,\bw}(\mu,\sigma^2,\varepsilon_{\overline{x}},\varepsilon_{\text{median}}|x_0,\m_3)$ with $\bw_k$ set to $1.6$, and repeated this run four times from overdispersed starting values to assess the convergence of the chains. Samples from the burn-in period were discarded. Figure\thinspace\ref{f:meanmedian2}A illustrates that our joint posterior error density remains virtually unchanged (compare to Figure\thinspace1C in  \cite{Ratmann2009e}). Next, we set $\alpha_0=2$ and $\beta_0=\bw_0=1000$ and ran \mcmcmyABC\ as above. Even though $\pi_\theta$ is now extremely broad, our joint posterior error density continues to identify model mismatch; see Figure\thinspace\ref{f:meanmedian2}B. 

ABC depends on the error threshold $\bw$, and so does \myABC. In order to identify model mismatch with $f_{\rho,\bw}(\beps|x_0,\m)$, existing conflicts among several summary statistics are uncovered by setting $\bw$ sufficiently small. For example, setting $\bw_k$ to $6.4$ such that the acceptance probability of \mcmcmyABC\ is larger than 80\%, the posterior error is very broad and does not suggest model mismatch; see Figure\thinspace\ref{f:meanmedian2}C.% Moreover, if $\bw_{\overline{x}}$ is very small and $\bw_{\text{median}}$ very large (or vice versa), then \myABC\ conditions effectively only on one summary and again the posterior error will not identify model mismatch.
\end{example}

In summary, the ability of \myABC\ to criticize a fitted model is strongly modulated by the choice of discrepancies and the error threshold $\bw$. Probing a model under particular assumptions on $\pi_\theta$ in the directions specified by $\beps$ is not guaranteed to uncover existing model mismatch. For example, using \myABC\ with the sample mean and the standard deviation (two independent summaries) in place of the sample mean and the median in Example\thinspace\ref{s:gaussmusigmapriors} fails to uncover existing model mismatch. Similarly, using the sample mean and the 25\% quantile fail to reveal model inconsistencies as clearly as the sample mean and the sample median. In principle, the contribution of \apostr{the data} to $f_{\rho,\bw}(\beps|x_0,\m)$ can only increase with larger $K$ and/or more stringent choices of $\bw$, and cannot be quantified by considering flat $\pi_{\beps}$ (see also A2).

In the directions determined by $\beps$ and $\bw$, the criticized model comprises the sampling model $\m$ and our prior assumptions $\pi_\theta$, and we acknowledge that \apostr{having no way to distinguish between prior and sampling model inadequacy is a difficulty} \cite{Robert2009b}. More work is needed here. 

\item {\it From an ABC perspective, using the posterior predictive $m(x|x_0,M)$ instead of the prior predictive $\pi(x|M)$ \apostr{requires same computing times} \cite{Robert2009b}.}
 
- In the context of ABC when the likelihood cannot be readily evaluated, the use of the posterior predictive (data) density
\begin{equation*}
m(x|x_0,\m)=\int f(x|\theta,\m)f(\theta|x_0,\m)\:dx
\end{equation*}
is complicated by the fact that samples from the true posterior density $f(\theta|x_0,\m)$ are in general not available. However, $m(x|x_0,\m)$ can be approximated by
\begin{equation*}
m_{\rho,\bw}(x|x_0,\m)=\int f(x|\theta,\m)f_{\rho,\bw}(\theta|x_0,\m)\:dx.
\end{equation*}
%which is sensitive to the choice of the prior $\pi_{\beps}$ and compound functions $x\to\rho_k\big(S_k(x),S_k(x_0)\big)$, questioning further the interpretability of derived quantities. 
An alternative approach for model criticism could be the approximate posterior predictive (APP) error density 
\begin{equation*}
L_{\rho,\bw,x_0}({\beps}|\m)=\int\delta\Big\{\Big(\rho_k\big(S_k(x),S_k(x_0)\big)=\varepsilon_k\Big)_{1:K}\Big\}\:m_{\rho,\bw}(x|x_0,\m)\:dx.
\end{equation*}
However, for a complex model the extra volatility induced by simulating from $f(\,\cdot\,|\theta,\m)$ means that re-simulations from $f_{\rho,\bw}(\theta|x_0,\m)$ need not meet the stringency requirement $\pi_{\beps}$. It might therefore also be useful to consider the weighted approximate posterior predictive (wAPP) error density
\begin{equation*}
f_{\rho,\bw,x_0}({\beps}|x_0,\m)\propto L_{\rho,\bw,x_0}({\beps}|\m)\pi({\beps}|\m).
\end{equation*}

Both $L_{\rho,\bw,x_0}({\beps}|\m)$ and $f_{\rho,\bw,x_0}({\beps}|x_0,\m)$ adopt a sequential approach to model criticism, comprising a training step (inference of $f_{\rho,\bw}(\theta|x_0,\m)$) and a testing step (APP or wAPP). The testing step adds a computational overhead to typical ABC procedures. For example, it takes about two minutes to evaluate our seven summaries on the {\it Saccharomyces cerevisiae} PPI data set \cite{Reguly2006}, and hence an extra $2\times 500 / 60\geq 16$hrs to obtain $500$ samples from APP on one computer. Assuming a large acceptance probability of 10\%, the extra time required to obtain $500$ samples from wAPP is more than $6$ days. By contrast, our posterior error incurs no additional computational cost because the discrepancies already computed in any ABC algorithm are only used to a fuller extent. Nevertheless, one might be prepared to pay this cost if the densities $L_{\rho,\bw,x_0}({\beps}|\m)$ and $f_{\rho,\bw,x_0}({\beps}|x_0,\m)$ would have an intrinsic advantage compared to our posterior error density $f_{\rho,\bw}({\beps}|x_0,\m)$.

In general, it is difficult to compare the behavior of our posterior error with APP and wAPP under model uncertainty. First, we note that $L_{\rho,\bw,x_0}({\beps}|\m)$ and our $f_{\rho,\bw}({\beps}|x_0,\m)$ are very different quantities, relating respectively to sequential and simultaneous approaches to model criticism. This is also reflected in their distinct asymptotic properties as $\bw\to 0$. Second, Example\thinspace\ref{ex:gaussian2paexample2b} in the Appendix demonstrates that, counter-intuitively, $f_{\rho,\bw}(\theta|x_0,\m)$ may be broader than $\pi_\theta$ under model uncertainty.

An additional complication to be considered with APP and wAPP is that the same aspects of the data are used to inform both the training and the testing phase. Hence, these quantities violate the fundamental requirement in statistical learning that the training data be independent from the testing data \cite{Hastie2001}. It is possible to use different aspects of the data during both stages, and this brings us back to the partially predictive and conditionally predictive densities previously discussed by Bayarri and Berger \cite{Bayarri1999}. Unfortunately, in real-world applications of ABC, it is often difficult to identify discrepancies that are independent of each other.

\item {\it Our estimator $\hat{\xi}$ to the augmented likelihood that is based on B repeat samples \apostr{cannot be used as a practical device because B is necessarily small} \cite{Robert2009PNAS} \apostr{$\dotsc$ in which case the non-parametric approximation is poor, or B is large in which case producing the x's is too time-consuming} \cite{Robert2009b}.}

- In our applications, we found that we obtained largest improvements in terms of the effective sampling size for small to moderate values of $B$ that are computationally feasible. Let us also recall that the choice of proposal kernel in $\beps$ is a crucial element of the second algorithm in \cite{Ratmann2009} and should not be omitted when considering its efficiency. We acknowledge that our observations may not readily extend to other applications. 

In the same way that we augmented standard ABC to what we call Std-\myABC\ in \cite{Ratmann2009}, it is straightforward to modify any existing ABC algorithm for the purpose of model criticism by using (i) many co-dependent, real-valued discrepancies and (ii) recording those discrepancies. For example, the Metropolis-Hastings sampler proposed by Marjoram et al. \cite{Marjoram2003} can be adapted to provide samples from the target distribution 
\begin{equation*}
\begin{split}
&f_{\rho,\bw}(d\theta,dx,d\beps|x_0,\m)=\quad\frac{\pi_\theta(\theta|\m)\pi_{\beps}(\beps|\m)f(x|\theta,\m)}{f_{\rho,\bw}(x_0|\m)}\:\big(\delta_{\rho_{1:K}(x)}(d\beps)\:dx\:d\theta\big),
\end{split}
\end{equation*}
where we put $\rho_{1:K}(x)=\big(\rho_k\big(S_k(x),S_k(x_0)\big)\big)_{1:K}$ for brevity. Here, $\delta_{\rho_{1:K}(x)}(d\beps)$ denotes the Dirac measure at the point $\rho_{1:K}(x)$.
Suppose initial values $\theta^0$, $x^0\sim f(\,\cdot\,|\theta^0,\m)$ and set $\varepsilon_{k}^0=\rho_k\big(S_k(x^0),S_k(x_0)\big)$.
\begin{singlespaceddescription}\label{s:mcmcabcmu}
\item[\mcmcmyABC 1] If now at $\theta$ propose a move to $\theta^\prime$ according to a proposal density $q(\theta\rightarrow\theta^\prime)$.
\item[\mcmcmyABC 2] Generate $x^\prime\sim f(\,\cdot\,|\theta^\prime,\m)$ and compute $\varepsilon_k^\prime=\rho_k\big(S_k(x^\prime),S_k(x_0)\big)$ for $k=1,\dotsc,K$.
\item[\mcmcmyABC 3] Accept $(\theta^\prime,x^\prime,\varepsilon^\prime_{1:K})$ with probability
\begin{equation*}
mh(\theta,x,\beps;\theta^\prime,x^\prime,\varepsilon^\prime_{1:K})=\quad\min\big\{1\:,\:r_{\text{vanilla}}(\theta,x,\beps; \theta^\prime, x^\prime, \beps[\prime])\big\}
\end{equation*}
where
\begin{equation*}
r_{\text{vanilla}}(\theta,x,\beps\:;\:\theta^\prime, x^\prime, \beps[\prime])=\quad\frac{\pi_\theta(\theta^\prime|\m)\:q(\theta^\prime\rightarrow\theta)\:\pi_{\beps}(\varepsilon^\prime_{1:K}|\m)}{\pi_\theta(\theta|\m)\:q(\theta\rightarrow\theta^\prime)\:\pi_{\beps}(\beps|\m)},
\end{equation*}
and otherwise stay at $(\theta,x,\beps)$. Then return to \mcmcmyABC 1.
\end{singlespaceddescription}
As is standard practice, the algorithm is run sufficiently long after a certain \apostr{burn-in} period, and samples from the burn-in period are discarded. It is not difficult to show that marginally in $(\theta,\beps)$, \mcmcmyABC\ provides samples from $f_{\rho,\bw}(\theta,\beps|x_0,\m)$ Eq.\thinspace\ref{e:jointposterior} for suitable proposal kernels $q(\theta\rightarrow\theta^\prime)$ under our regularity assumptions (A6).
%the contrary. I do not think this is true, but I did not yet look into this in detail. I suppose the largest gain of smoothing/gain to reduce the noise in the Metropolis-Hastings ratio occurs for small to moderate values of $B$. Maybe better to emphasize that the PNAS algorithm may control the volatility of the data-generating process and hence may be useful in circumstances when this volatility is large.  Crucially, any ABC algorithm can be straightforwardly modified to sample from the ABCmu target density.
\end{myenumerateresume}

\section{Model criticism and model comparison}
\begin{myenumerateresume}
\item {\it \apostr{The Bayesian foundations of \myABC\ are questionable: the consequences of rejecting a model are ignored by \myABC\ but include constructing another model} \cite{Robert2009PNAS} and \apostr{this leads to wonder about the gain compared with using the Bayes factor} \cite{Robert2009b}. Moreover, \apostr{the estimation of Bayes' factors is even faster} \cite{Robert2009PNAS} and \apostr{provides a different answer} \cite{Robert2009b}.}
%- BF faster than what? Certainly not computing our marginal error density.\\
%- TODO: what hypothesis are tested with BF, and how would that relate to our alternative-free posterior error? It does not seem possible to compare BF vs our marginal posterior in a meaningful way.

- To us, model criticism and model comparison are important and complementary aspects of statistical reasoning. Indeed, methods for model comparison attempt to choose between candidate models, even if all of them do not match the data in one or several aspects well.

ABC is very flexible in that arbitrary data-generating processes $\m$ can be analyzed without the need to compute the likelihood, so long as the evaluation of the summary statistics is computationally tractable. \myABC\ makes possible to evaluate at no extra computational cost whether a model matches the observed data in terms of a large set of summary statistics, and to obtain useful indications how a faulty model should be modified. In our work, we found that \myABC\ thus enables to iterate rapidly through the initial stages of model design to identify one or a suite of models which are in agreement with the data, even when the likelihood cannot be readily evaluated. We believe that the ability of \myABC\ to offer statistical rigor at this point is highly valuable to areas of modern science where complex models are now formulated to explain and agree with data collected across the traditional boundaries of disciplines. For example, in biology, we face a wealth of data that is hard to analyze in its entirety under current computer resources (e.g. molecular genetic data), or we have one intricate data set (e.g. molecular interaction networks), or we cross boundaries of biological organization (e.g. systems biology). 

The methods presented in Ratmann et al. \cite{Ratmann2009} do not address the problem of choosing a model from a suite of candidates. Model comparison when the likelihood cannot be readily evaluated is not the topic of \cite{Ratmann2009}, and has been introduced elsewhere \cite{Wilkinson2007, Beaumont2008, Fagundes2007, Toni2009, Grelaud2009}.

\begin{figure}[tbp]
	\centering
		\begin{minipage}[b]{0.8\textwidth}
		%\vspace{-1cm}
			\begin{minipage}[b]{0.48\textwidth}
				\begin{minipage}[b]{0.001\textwidth}
					{\bf A}\newline\vspace{8cm}

				\end{minipage}
				\begin{minipage}[b]{0.99\textwidth}
					\includegraphics[type=pdf,ext=.pdf,read=.pdf,width=\textwidth]{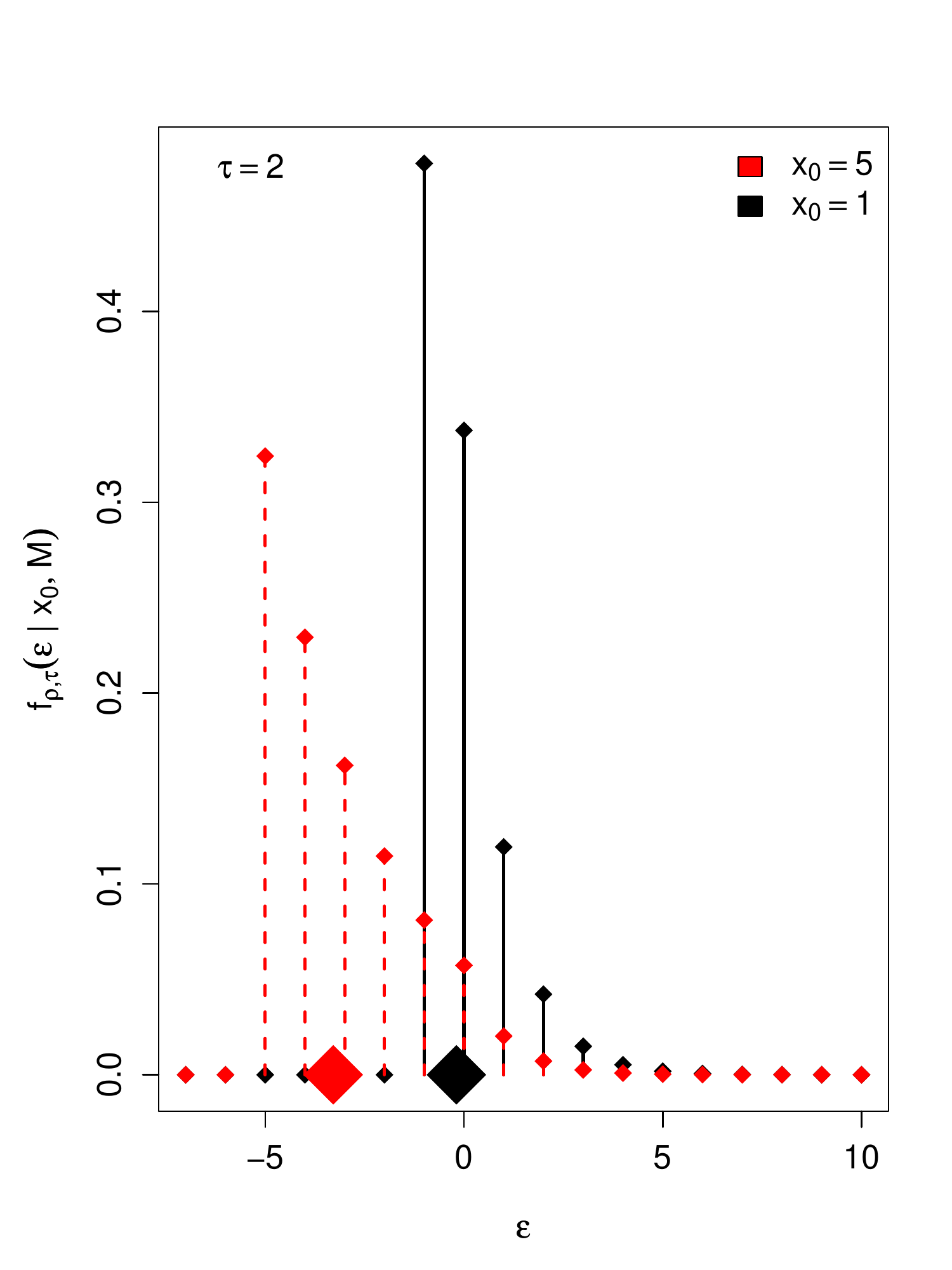}
				\end{minipage}
			\end{minipage}
			\begin{minipage}[b]{0.48\textwidth}
		%\vspace{-1cm}
				\begin{minipage}[b]{0.001\textwidth}
					{\bf B}\newline\vspace{8cm}

				\end{minipage}
				\begin{minipage}[b]{0.99\textwidth}
					\includegraphics[type=pdf,ext=.pdf,read=.pdf,width=\textwidth]{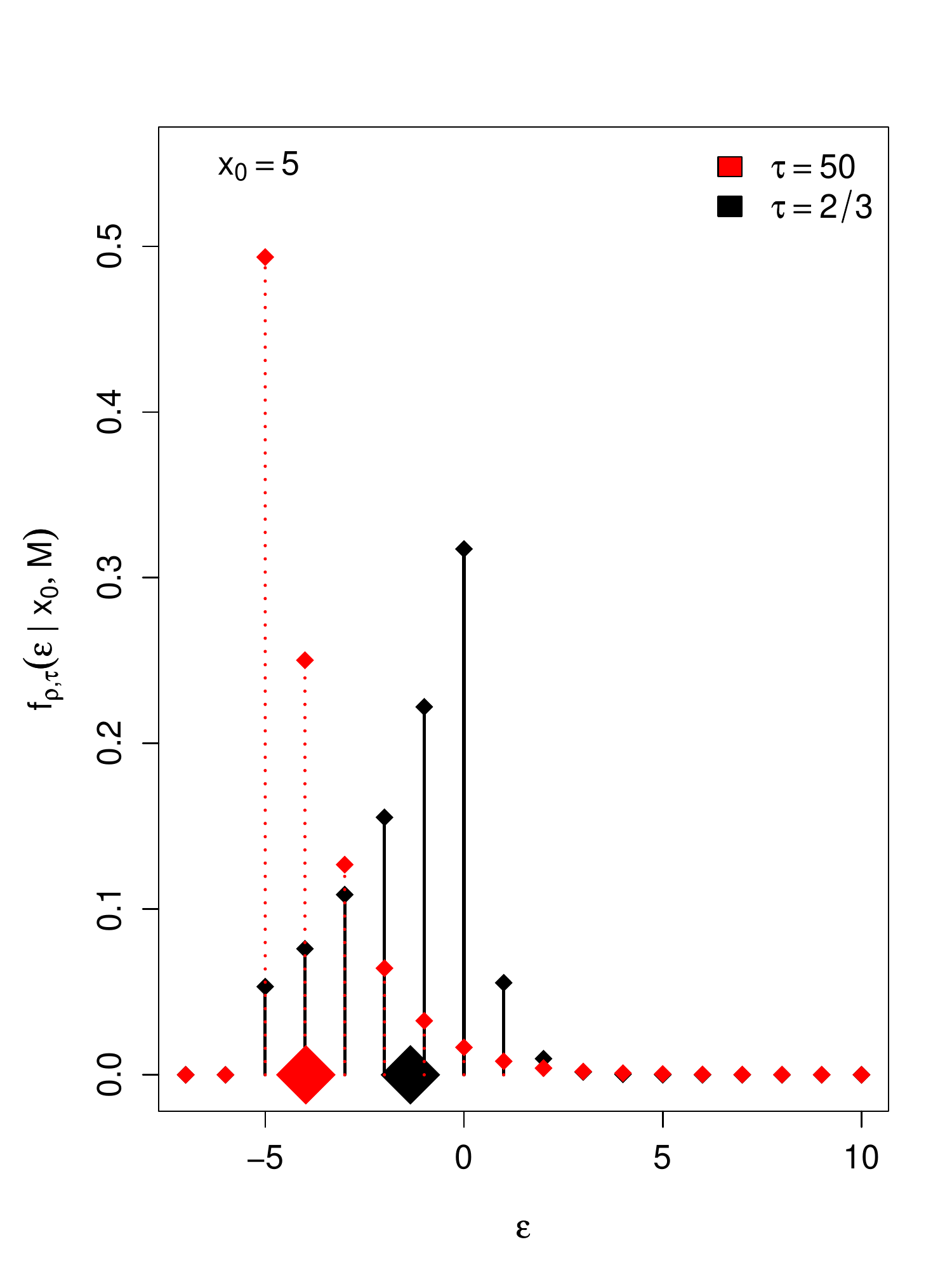}
				\end{minipage}
			\end{minipage}
		\end{minipage}
	\caption{Density plots of $f_{\rho,\bw}(\varepsilon|\m_1)$ in Example\thinspace\ref{s:poisson}. Posterior mean errors are indicated in large diamonds. (A) We fix $\bw=2$ and consider a data point $x_0=1$ that is in agreement with our prior belief $\pi_\theta=\text{Exp}(1)$ as well as a data point $x_0=5$ that differs from our prior model. In the latter case, the posterior mean error suggests mismatch between the model and the data. (B) We fix $x_0=5$ and consider the prior predictive error density (corresponding to an essentially flat $\pi_\varepsilon$ with $\bw=50$) and the posterior error density associated to $\bw=2/3$. Again, we observe that this fitted model is harder to criticize than the prior model.}\label{f:poisson1}
\end{figure}

\begin{figure}[tbp]
	\centering
		\begin{minipage}[b]{0.8\textwidth}
		%\vspace{-1cm}
			\begin{minipage}[b]{0.48\textwidth}
				\begin{minipage}[b]{0.001\textwidth}
					{\bf A}\newline\vspace{8cm}

				\end{minipage}
				\begin{minipage}[b]{0.99\textwidth}
					\includegraphics[type=pdf,ext=.pdf,read=.pdf,width=\textwidth]{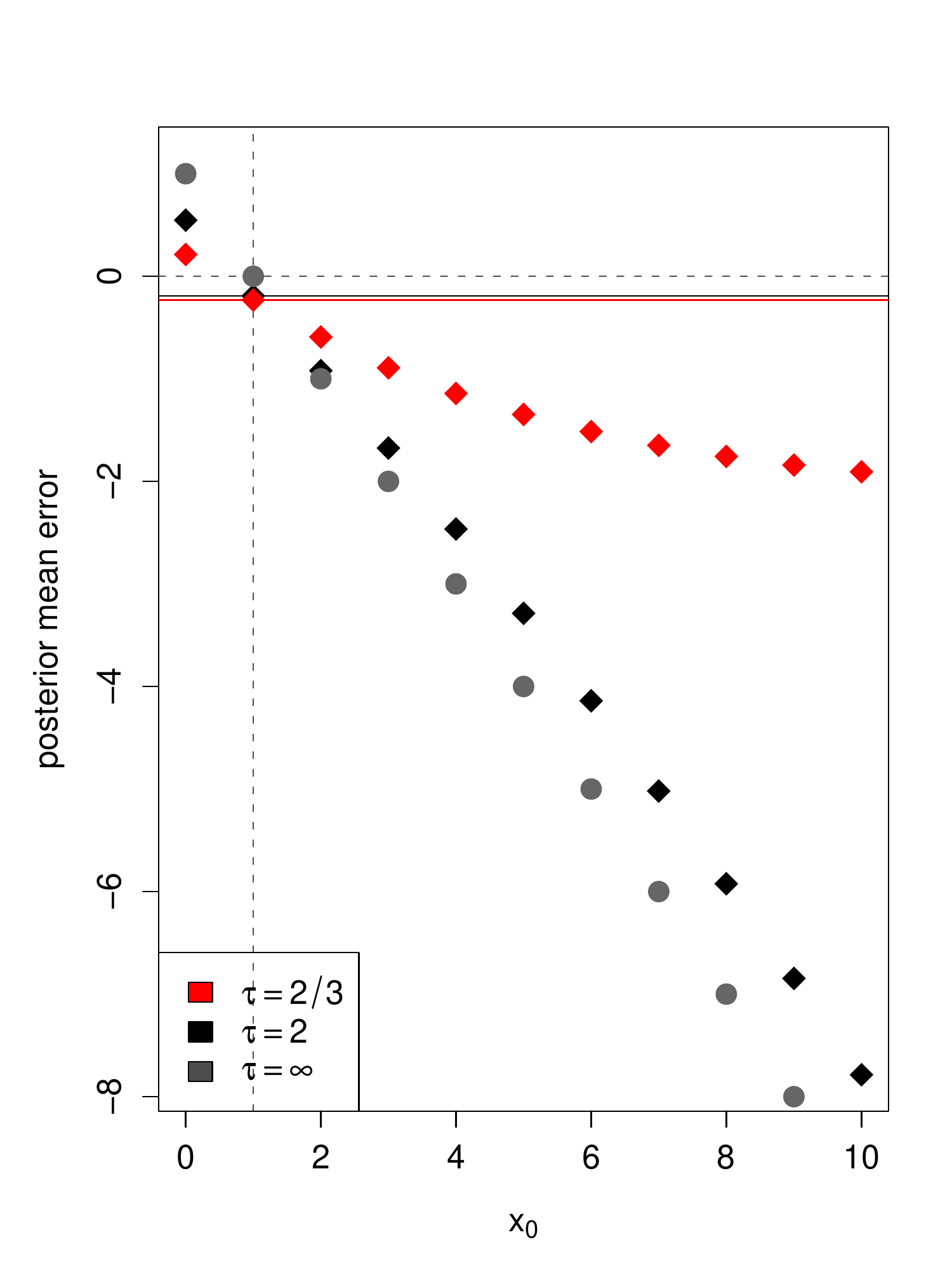}
				\end{minipage}
			\end{minipage}
			\begin{minipage}[b]{0.48\textwidth}
		%\vspace{-1cm}
				\begin{minipage}[b]{0.001\textwidth}
					{\bf B}\newline\vspace{8cm}

				\end{minipage}
				\begin{minipage}[b]{0.99\textwidth}
					\includegraphics[type=pdf,ext=.pdf,read=.pdf,width=\textwidth]{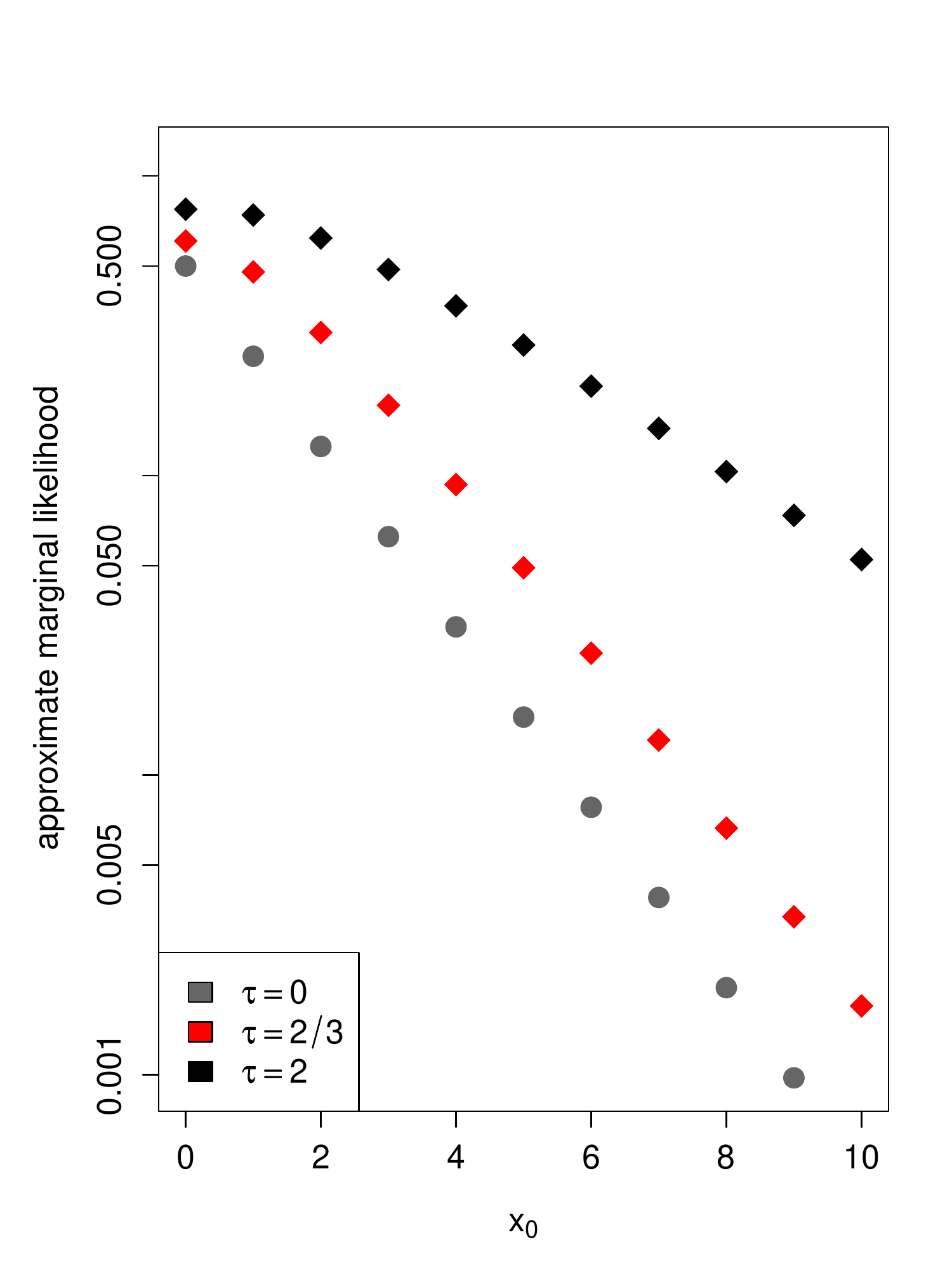}
				\end{minipage}
			\end{minipage}
		\end{minipage}
	\caption{(A) Plots of the posterior mean error $\int\varepsilon f_{\rho,\bw}(\varepsilon|x_0,\m_1)$ in Example\thinspace\ref{s:poisson} as a function of $x_0$ for $\bw=2/3$ (red), $\bw=2$ (black) and $\bw=\infty$ (grey). Provided the prior model is an adequate representation of the data ($x_0=1$), the prior predictive mean error is zero (dashed lines). By contrast, the posterior mean error is not zero in this case (red and black lines), simply because $L_\rho(\varepsilon|\m_1)$ is not symmetric. (B) Plots of the \myABC\ approximate marginal likelihood as a function of $x_0$ for $\bw=2/3$ (red), $\bw=2$ (black) and $\bw=\infty$ (grey). Note that this plot differs qualitatively from the one in \cite{Robert2009b} because RMC decided to truncate $\xi_{\theta,x_0}$ to positive values.}\label{f:poisson2}
\end{figure}

\begin{example}\label{s:poisson}
Let us re-visit RMC's Poisson example \cite{Robert2009PNAS, Robert2009b} in order to (a) illustrate model criticism with \myABC\ when the errors are discrete rather than continuous random variables, (b) inspect the case of asymmetric predictive error densities $L_\rho(\beps|\m)$ and (c) re-examine the behavior of the approximate marginal likelihood as presented in \cite{Robert2009b}, Figure 1.

Consider the Poisson model $\m_1$ of Example\thinspace\ref{ex:poisssonauglkl} and suppose that $\pi_\theta(\theta|\m_1)=\exp(-\theta)\Ind\big\{\theta\geq 0\big\}$. We have that
\begin{equation*}
\pi(x|\m_1)\quad=\quad\int_0^\infty\text{Poisson}(x;\theta)\exp(-\theta)\:d\theta\quad=\quad2^{-x-1}\Ind\{x\geq 0\},
\end{equation*}
and hence $L_\rho(\varepsilon|\m_2)\propto 2^{-x_0-\varepsilon-1}\Ind\{x_0+\varepsilon\geq 0\}$. The ABC kernel always depends on an \apostr{error threshold} $\bw$ (recall A1) and, given the form of  $L_\rho(\varepsilon|\m_2)$, we consider here $\pi_\varepsilon\colon\{0,1,-1,2,\dotsc\}\to\R^+_0$ with $\pi_\varepsilon(\varepsilon|\m_1)\propto2^{-\abs{\varepsilon}/\bw}$. Then, our marginal posterior error is
\begin{equation*}
f_{\rho,\bw}(\varepsilon|\m_1)\quad\propto\quad2^{-(x_0+\varepsilon+\abs{\varepsilon}/\bw+1)}\Ind\{x_0+\varepsilon\geq 0\},
\end{equation*}
with normalizing constant
\begin{equation}\label{e:poissonmlkl}
\begin{split}
&f_{\rho,\bw}(x_0|\m_1)\quad=\quad\sum_{\varepsilon=-x_0}^\infty2^{-(x_0+\varepsilon+\abs{\varepsilon}/\bw+1)}\\
&\quad=\quad2^{-(x_0+1)}\bigg(\Ind\{x_0>0\}\bigg[\frac{1-2^{(1-1/\bw)(x_0+1)}}{1-2^{1-1/\bw}}-1\bigg]\:+\:\frac{1}{1-2^{-1-1/\bw}}\bigg)
\end{split}
\end{equation}
under the assumption that $x_0\geq 0$. Figure\thinspace\ref{f:poisson1} illustrates the posterior error density for various choices of $x_0$ and $\bw$, and the respective posterior means are indicated in large diamonds. Note that $\pi_\varepsilon$ was here only chosen for reasons of analytical tractability, and we could still use our two-sided Exponential density $\pi_\varepsilon\colon\R\to\R^+_0$ where $\pi_\varepsilon(\varepsilon|\m)=1/\bw\exp\big(-2\abs{\varepsilon}/\bw\big)$, or the standard indicator function. Indeed, even if we do not know the set of possible discrete errors under a model $\m$, all we miss is the correct normalizing constant of $\pi_\varepsilon\colon\{0,1,-1,2,\dotsc\}\to\R^+_0$ where $\pi_\varepsilon(\varepsilon|\m)\propto\exp\big(-2\abs{\varepsilon}/\bw\big)$. This constant need not be known, see for example our algorithm \mcmcmyABC.

Figure\thinspace\ref{f:poisson2}A illustrates the posterior mean error $\int\varepsilon f_{\rho,\bw}(\varepsilon|\m_1)\:\varepsilon$ as a function of $x_0$ for various choices of $\bw$. Setting $\bw=\infty$, we obtain the prior predictive mean error, which is zero if the model corresponds well to the observed data ($x_0=1$). Since $L_\rho(\varepsilon|\m_2)$ is not symmetric around zero when the prior model is adequate (opposing A5), conditioning on error magnitude results in a slightly negative posterior mean error when $x_0=1$.

Let us recall that RMC decided to truncate the density $\varepsilon\to\xi_{\theta,x_0}(\varepsilon)$ to non-negative values, and then plotted the associated marginal likelihood $f_{\text{trunc}}(x_0|\m_1)=\iint \xi^{\text{trunc}}_{\theta,x_0}(\varepsilon)\pi_\theta(\theta|\m_1)\pi_\varepsilon(\varepsilon|\m_1)\:d\theta\:d\varepsilon$ as a function of $x_0$ in \cite{Robert2009b}, Figure 1. In Figure\thinspace\ref{f:poisson2}B, we plot the \myABC\ marginal likelihood Eq.\thinspace\ref{e:poissonmlkl} as a function of $x_0$. In this example, the approximate marginal likelihood decreases monotonically for all values of $\bw$, and only small values of $\bw$ provide a suitable approximation of the true marginal likelihood ($\bw=0$). Thus, the posterior mean error and the approximate marginal likelihood both depend on the precise value of the \apostr{error threshold}, suggesting that sensitivity analyses are required for \myABC\ as well as for complementary tools for model comparison that are based on approximate marginal likelihoods.
\end{example}

The Bayes' factor is a tool that may address both model comparison and model criticism, depending on the formulation of the null and alternative hypothesis. It is possible to devise approximate Bayes' factors to test the null hypothesis $\varepsilon=0$ versus the alternative $\varepsilon\neq 0$ as a surrogate measure for the hypothesis that the model describes the data adequately well, i.e. for the purpose of model criticism (unpublished results, but see \cite{Dickey1970, Verdinelli1995}). However, in both cases, the robustness of the Bayes' factor (with regard to the choice of $\bw$ and the quality of the numerical approximation of the ABC or \myABC\ target densities) is debatable (unpublished results). While we agree that computing the Bayes' factor proposed in \cite{Fagundes2007, Beaumont2008} is faster than computing posterior predictive checks, we also note that it cannot be faster than sampling from $f_{\rho,\bw}(\theta,\beps|x_0,\m)$ so long as the same ABC kernel (i.e. $\pi_{\beps}$) is used.

%- TODO: I have to understand the Poisson Bayes' factor; We are not surprised that methods for model comparison and model criticism yield different answers, as in the Poisson example. However, the Bayes' factor cannot be computed faster than our posterior error, because our posterior error does not incur any extra cost.

\item {\it \apostr{Comparing models via the posterior error is missing the model complexity penalisation from Bayesian model comparison} \cite{Robert2009PNAS}.}

- We acknowledge that model complexity is an important quantity to consider during model comparison. 

\item {\it \apostr{The choice of $\varepsilon$ and $\pi_\varepsilon(\varepsilon)$ is model dependent and the comparison [of models] reflects prior modeling, not data assessment} \cite{Robert2009PNAS}. Finally, \apostr{using the same $\bw$ across all models does not seem to be recommendable on a general basis} \cite{Robert2009PNAS}.}

- We agree that the choice of discrepancies (hence errors) and $\bw$ are application- and model specific. Although the same summaries can typically be used across models that attempt to explain the same data, model predictions will typically vary and hence the scales of the simulated summaries. This implies that the same $\bw$ may not always be used across different models. In this case, it may be difficult to compare the posterior error density $f_{\rho,\bw}(\varepsilon|x_0,\m)$ across different models. In \cite{Ratmann2009}, we only suggest to use $f_{\rho,\bw}(\varepsilon|x_0,\m)$ to compare each model against the observed data. 
\end{myenumerateresume}

\section{Conclusion}
We still find that \myABC\ enables us to comprehensively quantify discrepancies between a data-generating process $\m$ and the data, simultaneously with parameter inference even when the likelihood cannot be readily evaluated, thus providing valuable guidance on the interpretability of parameter estimates and on how to improve models. %We also believe that the ability to perform multi-dimensional model criticism with $K$-dimensional error densities such as  $f_{\rho,\bw}(\beps|x_0,\m)$ are valuable even when the likelihood can be readily computed. 

However, the method has its limitations. The posterior error reflects an interplay between the prior predictive error and the stringency with which that error is updated; recall Eq.\thinspace\ref{e:posteriorerror}. At present, there is no formal procedure to disentangle the contribution of $\pi_{\beps}$ and $L_{\rho}(\beps|\m)$ in $f_{\rho,\bw}(\beps|x_0,\m)$; this prompted us to caution that it is difficult to convincingly associate a formal, probabilistic framework with credibility intervals of $f_{\rho,\bw}(\varepsilon_k|x_0,\m)$ \cite{Ratmann2009}. In other words, there is no formal guarantee that zero is included in a 95\% credibility interval with a probability of 0.95 under the hypothesis that the prior model is correct. We agree that the methods proposed by Verdinelli and Wasserman \cite{Verdinelli1995} are promising for the purpose of model criticism via Bayes' factors, although the sharp hull hypothesis $\beps=0$ has limitations in itself \cite{Berger1987}. Moreover, we emphasize that $f_{\rho,\bw}(\beps|x_0,\m)$ cannot be thought of as a purely Bayesian quantity because $\pi_{\beps}$ also determines the approximation quality of $f_{\rho,\bw}(\theta|x_0,\m)$, recall Eq.\thinspace\ref{e:approxlkl}. In particular, this implies that the contribution of \apostr{the data} to $f_{\rho,\bw}(\beps|x_0,\m)$ cannot be directly quantified by setting $\pi_{\beps}$ uniform. Finally, we agree with RMC that the ABC and \myABC\ target densities, i.e.~$f_{\rho,\bw}(\theta|x_0,\m)$ and $f_{\rho,\bw}(\theta,\beps|x_0,\m)$, are sensitive to changes in the compound functions $x\to\rho_k\big(S_k(x),S_k(x_0)\big)$ (i.e. not invariant) and may attain different meanings under different choices of $\bw$. This leaves the whole method probabilistically coherent, but calls for sensitivity analyses.

Nonetheless, we believe that ABC and \myABC\ are useful to compare observed data and model simulations in a coherent way and to make inference on the model parameters as well as the error terms. It is difficult to understand  posterior quantities of $f_{\rho,\bw}(\theta|x_0,\m)$ in place of posterior quantities of the true posterior density $f(\theta|x_0,\m)$, but it makes good sense to interpret them as quantities that lie  between the prior and posterior density. With our inability to evaluate the likelihood $f(x_0|\theta,\m)$, we acknowledge that it is too cumbersome (or would take too long) to comprehend the data in full, and turn to those aspects $S_k$ of the data which we consider to be most relevant. Doing so, we retain some information of the available data and our posterior density $f_{\rho,\bw}(\theta|x_0,\m)$ updates our prior beliefs accordingly. Simultaneously, we can make use of the very same information to investigate the adequacy of a model $\m$ in explaining the data, and to update our prior predictions according to error magnitude. In conclusion, if we interpret $f_{\rho,\bw}(\theta|x_0,\m)$ as an update of our prior beliefs in $\theta$ and $f_{\rho,\bw}(\beps|x_0,\m)$ as an update of our prior predictive density under $\m$, then both are useful and meaningful quantities, particularly when the likelihood $f(\theta|x_0,\m)$ cannot be evaluated. %However, the relationship of the densities $f_{\rho,\bw}(\theta|x_0,\m)$, $f_{\rho,\bw}(\beps|x_0,\m)$ to the true posterior density $f(\theta|x_0,\m)$ and the true marginal likelihood $f(x_0|\m)$ remains often unclear.
%-We are interested to see future developments on nonparametric densities $\pi_{\beps}$ such that our quantities of interest $f_{\rho,\bw}(\theta|x_0,\m)$ and $f_{\rho,\bw}(\beps|x_0,\m)$ do not make repeat use of the same data. 

\subparagraph{Acknowledgements} We thank Julien Cornebise for insightful comments on an earlier version of this manuscript, as well as Christian Robert for insightful discussions that stimulated parts of these notes. OR gratefully acknowledges support from NSF grant NSF-EF-08-27416, CA is supported by an EPSRC Advance Research Fellowship, CW by the Danish Research Council, and S.R. by the Centre for Integrative Systems Biology at Imperial College as well as grant G-0600-609 from the Medical Research Council. 

\section{Appendix}

\begin{figure}[htbp]
	\centering
		\begin{minipage}[b]{0.8\textwidth}
		%\vspace{-1cm}
			\begin{minipage}[b]{0.48\textwidth}
				\begin{minipage}[b]{0.001\textwidth}
					{\bf A}\newline\vspace{9.5cm}

				\end{minipage}
				\begin{minipage}[b]{0.99\textwidth}
					\includegraphics[type=pdf,ext=.pdf,read=.pdf,width=\textwidth]{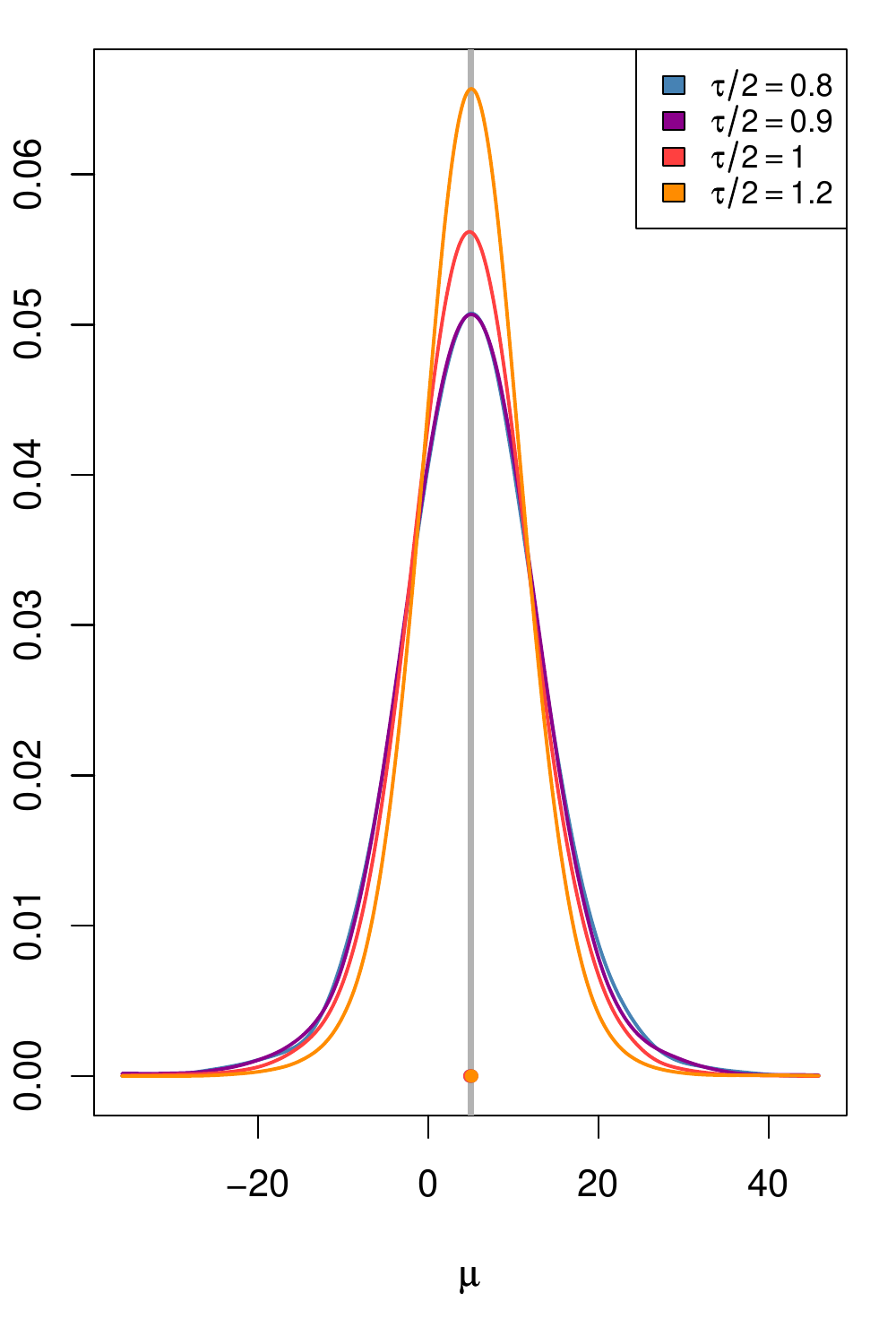}
				\end{minipage}
			\end{minipage}
			\begin{minipage}[b]{0.48\textwidth}
		%\vspace{-1cm}
				\begin{minipage}[b]{0.001\textwidth}
					{\bf B}\newline\vspace{9.5cm}

				\end{minipage}
				\begin{minipage}[b]{0.99\textwidth}
					\includegraphics[type=pdf,ext=.pdf,read=.pdf,width=\textwidth]{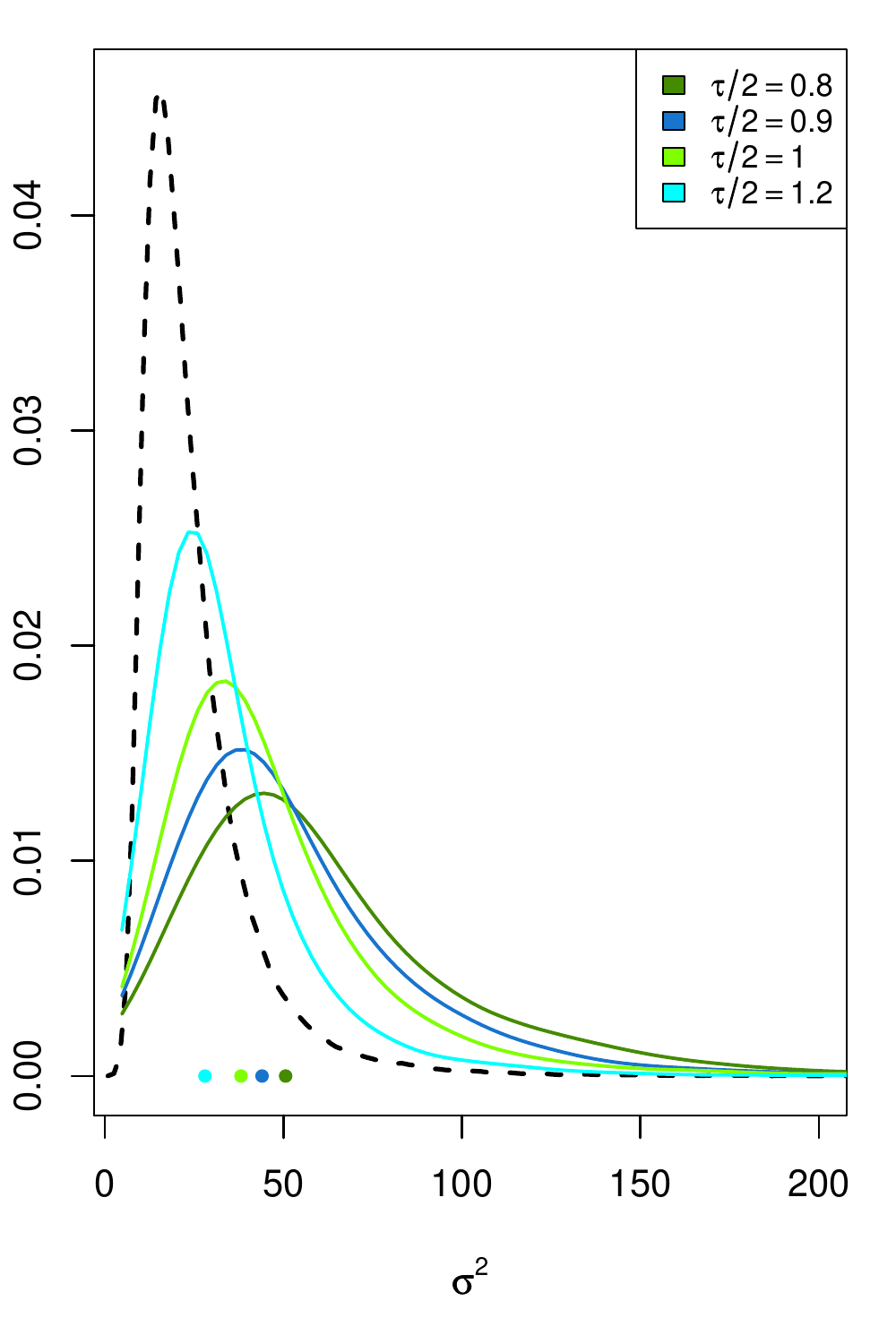}
				\end{minipage}
			\end{minipage}
		\end{minipage}
	\caption{Numerical estimates of (A) $f_{\rho,\bw}(\mu|x_0,\m_3)$ and (B) $f_{\rho,\bw}(\sigma^2|x_0,\m_3)$ in Example\thinspace\ref{ex:gaussian2paexample2b} for decreasing values of $\bw_{\SYMMm}$ (different colors). The respective marginal prior densities are overlaid (black, dashed).}\label{f:gaussian2pasymm1b}
\end{figure}

\begin{example}\label{ex:gaussian2paexample2b} 
Reconsider the data set $x_0$ of $n=100$ independent samples that are Exponentially distributed with rate $1/\mu_t=0.2$. We believe again that each sample of $x_0$ is generated from $\N(\mu,\sigma^2)$ with $\mu\in\R$, $\sigma^2\geq 0$ unknown, and take a conjugate normal inverse-gamma prior density for $\mu$ and $\sigma^2$ with hyperparameters $\mu_0\in\R$, $n_0=1$, $\alpha_0>0$ and $\beta_0>0$
\begin{equation*}
\begin{split}
\pi(\sigma^2|\m_3)&= f_{\text{IG}(\alpha_0,\beta_0)}(\sigma^2)=\quad\beta_0^{\alpha_0}\big(\sigma^{-2}\big)^{\alpha_0+1}\exp\big(-\beta_0\sigma^{-2}\big)\:\big/\:\Gamma(\alpha_0)\\
\pi(\mu|\sigma^2,\m_3)&=f_{\mathcal{N}(\mu_0,\sigma^2/n_0)}(\mu)\\[2mm]
\pi_{\beps}(\beps|\m_3)&=\prod_{k=1}^K1/\bw_k\Ind\big\{\bigabs{\varepsilon_k}\leq\bw_k/2\big\},
\end{split}
\end{equation*}
with hyperparameters set to $\mu_0= 5$, $n_0=1$, $\alpha_0=4$ and $\beta_0=75$. We ran Std-\myABC\ based on the summary $\SYMMm(x)=\overline{x}-\text{median}(x)$, $\rho\big(\Sset(x),\Sset(x_0)\big)=\SYMMm(x)-\SYMMm(x_0)$ and the above conjugate prior in $\theta$ for 20$,$000 iterations to obtain samples from the joint posterior density $f_{\rho,\bw}(\mu,\sigma^2,\varepsilon_{\SYMMm}|x_0,\m_3)$ for various values of $\bw$. Interestingly, the approximate posterior density $f_{\rho,\bw}(\theta|x_0,\m)$ broadens for decreasing values of $\bw$, as shown in Figures\thinspace\ref{f:gaussian2pasymm1b}(A-B). %Note that the prior density Eq.\thinspace\ref{se:IGconjugateprior} introduces dependencies between $\mu$ and $\sigma^2$ that are propagated into $f_{\rho,\bw}(\mu|x_0,\m_3)$. % illustrating the fact that it is difficult to perform model checking on the basis of quantities of $f_{\rho,\bw}(\theta|x_0,\m_3)$ which are themselves convoluted functions of the likelihood model under question.
\end{example}

%%%%%%%%%%%%%%%%%%%%%%%%%%%%%%%%%%%%%%%%%%%%%%%%%%%%%%%%%%%%%%%%
\begin{small}

\end{small}
%%%%%%%%%%%%%%%%%%%%%%%%%%%%%%%%%%%%%%%%%%%%%%%%%%%%%%%%%%%%%%%%
\end{document}